\documentclass{article}
\PassOptionsToPackage{numbers,compress,sort&compress}{natbib}
\usepackage[main, preprint]{neurips_2026}
\usepackage[utf8]{inputenc}
\usepackage[T1]{fontenc}
\usepackage[colorlinks=true,allcolors=blue,breaklinks=true]{hyperref}
\usepackage{url}
\usepackage{booktabs}
\usepackage{amsfonts}
\usepackage{nicefrac}
\usepackage{microtype}
\usepackage{xcolor}
\usepackage{amsmath,amssymb}
\usepackage{graphicx}
\usepackage{tabularx,array,longtable,threeparttable,multirow,makecell}
\usepackage{enumitem}
\usepackage[nameinlink,noabbrev]{cleveref}
\usepackage{subcaption}
\usepackage{float}
\usepackage{flafter}
\usepackage{placeins}
\usepackage{etoolbox}
\usepackage{listings}
\usepackage{pdfpages}
\usepackage{pifont}
\usepackage{algorithm}
\usepackage{algpseudocode}
\usepackage[most]{tcolorbox}
\tcbuselibrary{skins,breakable}
\usepackage{xspace}
\usepackage[percent]{overpic}
\usepackage{pdfpages}

\newcommand{\cmark}{\textcolor{green!60!black}{\ding{51}}}
\newcommand{\xmark}{\textcolor{red!75!black}{\ding{55}}}
\definecolor{promptblue}{HTML}{2E2EBB}
\newtcblisting{promptbox}[2][promptblue]{
  colback=white,
  colframe=#1,
  colbacktitle=#1!10,
  coltitle=black,
  title=#2,
  fonttitle=\bfseries\sffamily,
  listing only,
  listing options={basicstyle=\ttfamily\footnotesize,breaklines=true},
  breakable,
  enhanced,
  boxrule=0.8pt,
  top=2pt,bottom=2pt,left=2pt,right=2pt,
  before upper=\noindent,
}
\newtcblisting{promptboxsmall}[2][promptblue]{
  colback=white,
  colframe=#1,
  colbacktitle=#1!10,
  coltitle=black,
  title=#2,
  fonttitle=\bfseries\sffamily,
  listing only,
  listing options={basicstyle=\ttfamily\scriptsize,breaklines=true},
  breakable,
  enhanced,
  boxrule=0.8pt,
  top=2pt,bottom=2pt,left=2pt,right=2pt,
  before upper=\noindent,
}
\lstdefinestyle{pseudocode}{basicstyle=\ttfamily\small,breaklines=true,columns=fullflexible,frame=single,showstringspaces=false,keepspaces=true}
\lstdefinestyle{promptfile}{basicstyle=\ttfamily\scriptsize,breaklines=true,columns=fullflexible,frame=single,showstringspaces=false,keepspaces=true}
\newcolumntype{Y}{>{\raggedright\arraybackslash}X}
\newcommand{\framework}{\emph{AI CFD Scientist\xspace}}
\newcommand{\foamagent}{Foam-Agent\xspace}

\graphicspath{{figs/}}
\pretocmd{\subsection}{\FloatBarrier}{}{}

\title{\framework: Toward Open-Ended Computational Fluid Dynamics Discovery with Physics-Aware AI Agents}
\date{}

\author{%
  Nithin Somasekharan$^1$ 
  \quad 
  Rabi Pathak$^1$ 
  \quad 
  Manushri Dhanakoti$^1$ 
  \quad 
  Tingwen Zhang$^1$ 
  \\ 
  \textbf{Ling Yue}$^1$ 
  \quad 
  \textbf{Andy Zhu}$^1$ 
  \quad 
  \textbf{Shaowu Pan}$^1$\thanks{Corresponding author: \texttt{pans2@rpi.edu}}
  \\\\
  $^1$Rensselaer Polytechnic Institute
}

\begin{document}
\maketitle
\begin{abstract}

{Recent LLM-based agents have closed substantial portions of the scientific discovery loop in software-only machine-learning research, in chemistry, and in biology. Extending the same loop to high-fidelity physical simulators is harder, because solver completion does not imply physical validity and many failure modes appear only in field-level imagery rather than in solver logs. We present \framework, an open-source AI scientist for computational fluid dynamics (CFD) that, to our knowledge, is the first to span literature-grounded ideation, validated execution, vision-based physics verification, source-code modification, and figure-grounded writing within a single inspectable workflow. Three coupled pathways cover parameter sweeps within a fixed solver, case-local C\texttt{++} library compilation for new physical models, and open-ended hypothesis search against a reference comparator, all running on OpenFOAM through \foamagent. At the center of the framework is a vision-language physics-verification gate that inspects rendered flow fields before any result is accepted, rerun, or written into a manuscript. On five tasks under a shared GPT-5.5 backbone, {\framework} autonomously discovers a Spalart--Allmaras runtime correction that reduces lower-wall $C_f$ RMSE against DNS by $7.89\%$ on the periodic hill at $Re_h{=}5600$; under matched LLM cost, two strong general AI-scientist baselines (ARIS, DeepScientist) execute partial CFD workflows but lack the domain-specific validity gates needed to convert runs into defensible scientific claims; and a controlled planted-failure ablation shows that the vision-language gate detects $14$ of $16$ silent failures missed by solver-level checks. Code, prompts, and run artifacts are released at \url{https://github.com/csml-rpi/cfd-scientist}.}
\end{abstract}

\section{Introduction}
\label{sec:intro}

{Large language model agents have closed substantial portions of the scientific discovery loop in software-only machine-learning research~\citep{lu2024aiscientist,yamada2025aiscientistv2}, in chemistry~\citep{bran2024chemcrow}, and in biology~\citep{huang2024crisprgpt}. Extending these systems to physical sciences whose evidence comes from high-fidelity simulators is the next frontier and remains underexplored, in part because the discovery loop interacts with the simulator at a level deeper than text-mediated tool use.}

{Computational fluid dynamics (CFD) makes this loop particularly strict for three reasons. First, solver completion does not imply physical validity: a case can run cleanly while still using the wrong geometry, missing a key flow feature, or producing degenerate output. These failure modes are typically invisible to solver logs.\footnote{For example, a backward-facing-step case can converge cleanly while a reattachment-length extractor returns a wrong-sign value: invisible in the solver log, but obvious in a $C_f$ plot.} Second, validity gates are themselves scientific objects: mesh independence and reference-data alignment must be confirmed before any claim, not assumed. Third, the closure model is a research variable, edited at the C\texttt{++} level rather than swapped in a config, so source-code modification is part of the hypothesis space rather than a configuration option.}

{Two lines of work approach this loop from opposite sides but neither covers it end-to-end. Generic AI-scientist frameworks~\citep{yamada2025aiscientistv2,schmidgall2025agentlab,yang2026aris,weng2026deepscientist} automate ideation, code, plotting, and writing, but they were designed for software-only ML workflows and lack the physical-validity gates that distinguish a runnable simulation from a defensible scientific claim. CFD-specific agents~\citep{yue2025foamagent,chen2025metaopenfoam2,xiao2026flamepilot,turbulenceai} automate case setup, execution, and parts of post-processing on OpenFOAM-style substrates, but stop short of the full discovery loop. The closest related system, \emph{turbulence.ai}~\citep{turbulenceai}, frames an AI scientist for fluid mechanics that formulates ideas, orchestrates experiments, and drafts reports, yet remains closed-source and, based on public documentation as of submission, does not expose a vision-language physics-verification gate, a mesh-independence gate, or open-ended source-level discovery as first-class subsystems.}

{We present \framework, an open-source AI scientist for CFD that, to our knowledge, is the first to span literature-grounded ideation, validated execution, vision-based physics verification, source-code modification, and figure-grounded writing within a single inspectable workflow. The framework runs on OpenFOAM through \foamagent~\citep{yue2025foamagent} and exposes three coupled pathways: regular experimentation through parameter sweeps within a fixed solver, source-code modification that compiles case-local C\texttt{++} libraries for new physical models, and open-ended hypothesis search that autonomously edits source code and coefficients against a reference comparator. At the center of the framework is a vision-language physics-verification gate that inspects rendered flow fields before any result is accepted, rerun, or written into a manuscript: a subsystem absent from the AI-scientist baselines we compare against. The architecture follows five operational design principles distilled from CFD practice, detailed in \cref{sec:cfd_scientist}.}

\begin{figure}[h]
\centering
\includegraphics[width=\textwidth]{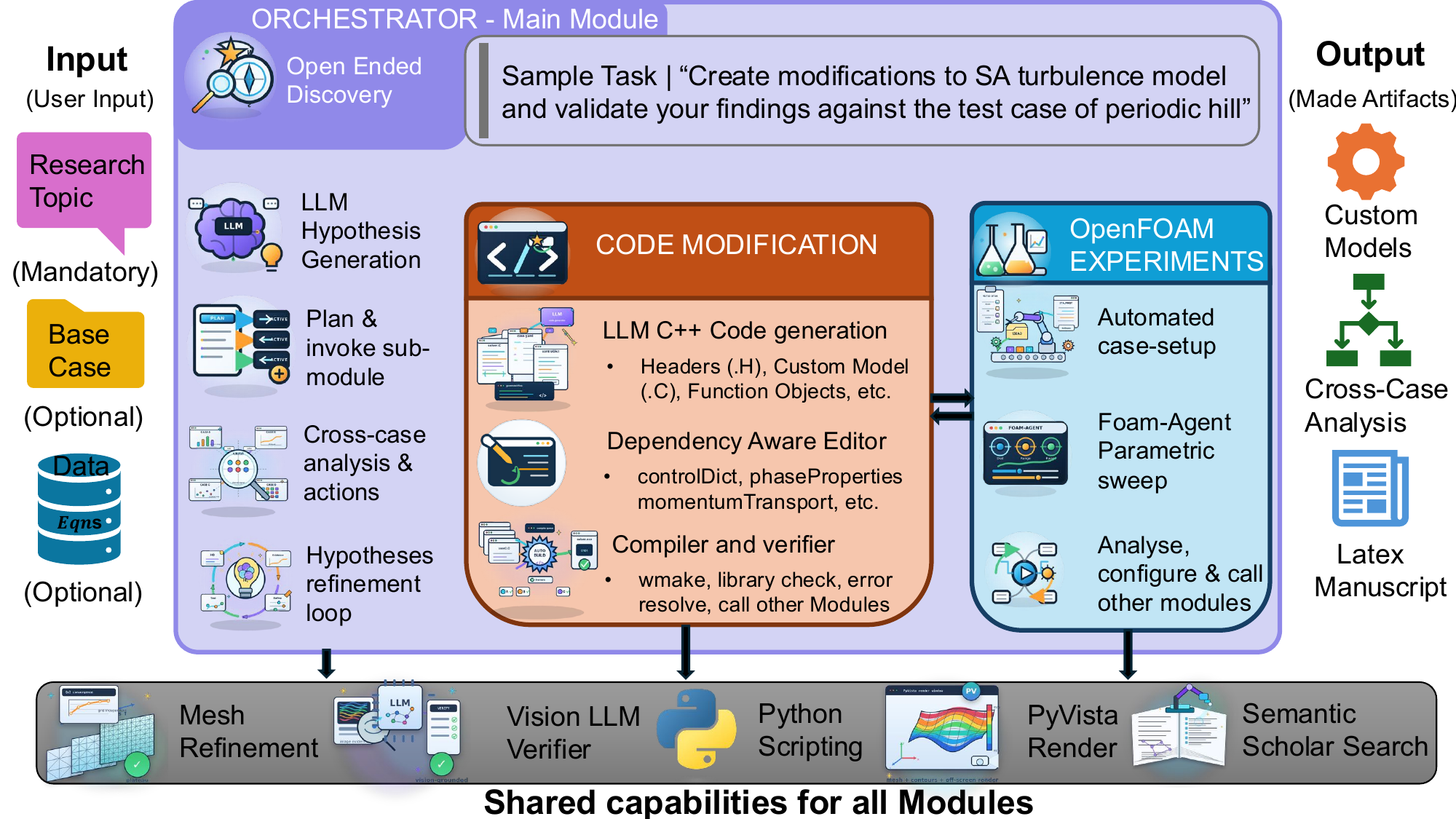}
\caption{Architecture of \framework. A natural-language topic, optional base case, and optional reference data is passed as input to the framework. Three first-class pathways execute under a shared capability bus: (i) \emph{regular experimentation} via literature-aware ideation, requirement validation, mesh-independence gating, and \foamagent execution; (ii) \emph{code modification} that patches and compiles case-local C\texttt{++} model libraries; (iii) \emph{open-ended discovery} that wraps both modules in an outer hypothesis loop. A VLM physics gate inspects rendered flow fields before any result is accepted, rerun, or written.}
\label{fig:system_overview}
\end{figure}

{On five tasks under a shared GPT-5.5 backbone, \framework executes regular experimentation, custom-model compilation, and open-ended discovery; in the open-ended task, the system autonomously discovers a Spalart--Allmaras runtime correction that reduces lower-wall $C_f$ RMSE against DNS by $7.89\%$ on the periodic hill at $Re_h{=}5600$. Under matched LLM cost, two strong general AI-scientist baselines (ARIS~\citep{yang2026aris}, DeepScientist~\citep{weng2026deepscientist}) execute partial CFD workflows but lack the domain-specific validity gates needed to convert runs into defensible scientific claims. A controlled planted-failure ablation shows that the vision-language physics gate detects $14$ of $16$ silent failures missed by solver-level checks. 

\begin{table}[h]
\centering
\caption{Positioning \framework{} against generic AI-scientist frameworks and CFD-specific agents.}
\label{tab:positioning}
\scriptsize
\setlength{\tabcolsep}{2.0pt}
\renewcommand{\arraystretch}{1.14}

\newcommand{\poshead}[1]{%
\parbox[c][4.8em][c]{\linewidth}{\centering\bfseries #1}%
}
\newcommand{\possyshead}{%
\parbox[c][4.8em][c]{3.65cm}{\raggedright\bfseries System}%
}

\begin{tabularx}{\textwidth}{
>{\raggedright\arraybackslash}p{3.65cm}
*{8}{>{\centering\arraybackslash}X}
}
\toprule
\possyshead
& \poshead{Literature\\Survey}
& \poshead{Novelty\\Filtering}
& \poshead{CFD\\Execution}
& \poshead{Mesh\\Independence\\Study}
& \poshead{Simulator\\Source Code\\Editing}
& \poshead{VLM-Based\\Physics\\Check}
& \poshead{Paper\\Generation}
& \poshead{Reference\\Data\\Ingestion} \\
\midrule

\multicolumn{9}{l}{\textit{Generic AI-scientist frameworks designed primarily for ML research}}\\
\midrule

AI Scientist (-v2)\citep{yamada2025aiscientistv2}
& \cmark & \xmark & \xmark & \xmark & \xmark & \xmark & \cmark & \xmark \\

Agent Laboratory\citep{schmidgall2025agentlab} / AgentRxiv\citep{schmidgall2025agentrxiv}
& \cmark & \xmark & \xmark & \xmark & \xmark & \xmark & \cmark & \xmark \\

AI co-scientist\citep{gottweis2025couscientist}
& \cmark & \xmark & \xmark & \xmark & \xmark & \xmark & \xmark & \xmark \\

CycleResearcher\citep{weng2025cycleresearcher}
& \cmark & \xmark & \xmark & \xmark & \xmark & \xmark & \cmark & \xmark \\

DeepScientist\citealp{weng2026deepscientist}
& \xmark & \xmark & \cmark & \xmark & \xmark & \xmark & \xmark & \xmark \\

ARIS\citep{yang2026aris}
& \xmark & \xmark & \cmark & \xmark & \cmark & \xmark & \xmark & \xmark \\

\midrule
\multicolumn{9}{l}{\textit{CFD-specific agents}}\\
\midrule

MetaOpenFOAM\citep{chen2024metaopenfoam}, ChatCFD\citep{fan2025chatcfd}, OpenFOAMGPT\citep{pandey2025openfoamgpt}
& \xmark & \xmark & \cmark & \xmark & \xmark & \xmark & \xmark & \xmark \\

\foamagent\citep{yue2025foamagent}
& \xmark & \xmark & \cmark & \xmark & \xmark & \xmark & \xmark & \xmark \\

CFDagent\citep{xu2025cfdagent} / SwarmFoam\citep{yang2026swarmfoam} / PhyNiKCE\citep{fan2026phynikce} / CFD-copilot\citep{dong2025cfdcopilot}
& \xmark & \xmark & \cmark & \xmark & \xmark & \xmark & \xmark & \xmark \\

\textit{turbulence.ai}\citep{turbulenceai}
& \cmark & \xmark & \cmark & \xmark & \xmark & \xmark & \cmark & \xmark \\

FlamePilot\citep{xiao2026flamepilot}
& \cmark & \xmark & \xmark & \xmark & \xmark & \xmark & \xmark & \xmark \\

\midrule
\framework{} \textbf{(this work)}
& \cmark & \cmark & \cmark & \cmark & \cmark & \cmark & \cmark & \cmark \\
\bottomrule

\end{tabularx}
\end{table}

\section{Related Work}
\label{sec:related}
\paragraph{Robot scientists and autonomous laboratories.} Closing the scientific loop predates LLMs. The \emph{Robot Scientist} systems \citep{king2009automation,sparkes2010robotscientists} demonstrated end-to-end hypothesis generation and physical experimentation in molecular biology, and symbolic-regression engines such as Eureqa \citep{schmidt2009distilling} automated equation discovery from data. More recent self-driving laboratories \citep{gongora2020bear,macleod2022pareto,seifrid2022selfdrivinglab} fuse robotic experimentation with Bayesian-optimization planners. These systems target chemistry, materials, and biology, where ground truth comes from physical measurement; they do not transfer to CFD, where validity depends on closure choices, mesh resolution, and physical interpretation of computed fields rather than wet-lab readouts.

\paragraph{LLM-based AI-scientist frameworks.} A second wave of systems closes the same loop in pure software using LLMs. \emph{The AI Scientist} and \emph{AI Scientist-v2} \citep{lu2024aiscientist,yamada2025aiscientistv2} produce end-to-end ML papers from a research idea; \emph{Agent Laboratory} and \emph{AgentRxiv} \citep{schmidgall2025agentlab,schmidgall2025agentrxiv} formalize multi-agent collaboration and inter-paper memory; \emph{AI co-scientist} \citep{gottweis2025couscientist} layers critique-driven refinement. \emph{CycleResearcher}, \emph{AI-Researcher}, and \emph{Zochi} \citep{weng2025cycleresearcher,tang2025airesearcher,intology2025zochi} emphasize iterative refinement and tool-use; \emph{DeepScientist} \citep{weng2026deepscientist} and \emph{ARIS} \citep{yang2026aris} are the most recent strong baselines, both built around long-context execution loops, and are the two systems used in our head-to-head comparison. Domain instances exist in chemistry and biology, example: \emph{ChemCrow}, autonomous chemistry agents, and CRISPR-GPT \citep{bran2024chemcrow,boiko2023autonomouschem,huang2024crisprgpt}. Evaluation infrastructure (Bohrium--SciMaster, AstaBench, PaperBench, MLR-Bench \citep{zhang2025bohriumscimaster,darcy2025astabench,starace2025paperbench,chen2025mlrbench}) scores artifact quality on ML research workflows. 

\paragraph{CFD- and OpenFOAM-specific agents.} A parallel line of work targets CFD itself. PythonFOAM and foamlib \citep{maulik2022pythonfoam,gerlero2025foamlib} expanded the Python surface for case manipulation and in-situ analysis. LLM-centered systems then moved from prompt assistance to structured orchestration: FoamPilot \citep{xu2024foampilot} and AutoCFD \citep{dong2025autocfd} are early prompt-driven assistants, OpenFOAMGPT and MetaOpenFOAM (with optimized variants) \citep{pandey2025openfoamgpt,chen2024metaopenfoam,chen2025metaopenfoam2,chen2025optmetaopenfoam,feng2025openfoamgpt2} structure the case-authoring workflow, and \foamagent \citep{yue2025foamagent} adds RAG-based retrieval and a reviewer loop. ChatCFD \citep{fan2025chatcfd}, CFDagent \citep{xu2025cfdagent}, SwarmFoam \citep{yang2026swarmfoam}, PhyNiKCE \citep{fan2026phynikce}, CFD-copilot \citep{dong2025cfdcopilot}, \textit{turbulence.ai} \citep{turbulenceai}, and FlamePilot \citep{xiao2026flamepilot} extend the surface to chat-driven workflows, multi-agent decomposition, physics constraints, and combustion. General coding agents also solve a subset of OpenFOAM workflows by reusing tutorials \citep{xiao2026codingagentscfd}, and a separate line asks whether LLMs can act as neural fluid surrogates \citep{chen2026llm4fluid}. None of these systems combine all the relevant features needed for automating CFD discovery. This gap motivates \framework{}.
\section{CFD Scientist}
\label{sec:cfd_scientist}

\framework{} encodes CFD discovery as a set of expert-written prompts, guidelines, and execution pathways rather than a generic chat loop. We provide two implementations: a checkpointed LangGraph workflow for end-to-end orchestration, and a modular skills-based version whose components can be reused inside other orchestrators. In both forms, agents exchange structured artifacts such as study JSON, requirement paragraphs, source-edit plans, run directories, figure manifests, interpretation JSON, and manuscript drafts as shown in \cref{fig:system_overview}. The design follows five principles distilled from CFD practice: \textbf{(P1)} physical validity is not log-readable, so image-level inspection is mandatory; \textbf{(P2)} source code modification is a research object rather than a configuration option; \textbf{(P3)} mesh independence is a required convergence gate; \textbf{(P4)} agents must not hallucinate an alternate experiment, swap the swept variable, or relax success criteria in order to make a failing case easier to run; \textbf{(P5)} every claim in the generated manuscript must trace back to a specific figure, numerical value, or interpretation record produced by a case that passed its validity gates, never to the model's prior knowledge.
\paragraph{Three pathways.} \emph{Regular experimentation:} This pathway runs CFD simulation studies without modifying simulator source code. Given a research topic, the literature-aware ideation agent retrieves Semantic Scholar records, synthesizes candidate gaps, and emits a structured study JSON. A string-similarity novelty filter rejects near-duplicate ideas and triggers re-prompting when needed. The specification agent then converts each experiment into a single-paragraph requirement. A validator checks solver availability, time-control consistency, boundary-condition completeness, and unit consistency; failed specifications are rewritten through a repair prompt. Validated requirements are passed to \foamagent{}~\citep{yue2025foamagent}, which generates the case dictionaries, executes the simulation, and performs low-level error correction. \emph{Code modification:} for studies that require a model not present in the OpenFOAM source code, an expert-written code-mod agent generates C\texttt{++} source and dictionary edits, compiles a \emph{case-local} library under \texttt{\{case\}/customModels/}, and uses compiler diagnostics as structured feedback; a smoke test verifies the library loads and produces interpretable fields before any sweep. \emph{Open-ended discovery:} given an abstract goal such as \emph{find a novel turbulence-model modification that better matches a given DNS reference}, or any user-supplied objective with a comparator, an outer hypothesis loop autonomously generates and tests candidate ideas without further human input. At each iteration it proposes a concrete edit (a source-code change to the turbulence model, a coefficient or parameter adjustment, or a new diagnostic script), invokes the code-modification and regular-experimentation pathways to compile and run it as a real OpenFOAM case, and compares the resulting flow field against both the reference data and the unmodified baseline. Iterations are scored by a user-specified comparator, checkpointed  and promoted only when the score improves over baseline.

\paragraph{Mesh-independence gate.} A baseline mesh is selected from a starter case, literature, or generated by \foamagent. A refined mesh is constructed with $\sim$10\% near-wall and $\sim$5\% bulk refinement, preserving topology, blocking, and meshing method. Baseline and refined cases run with identical models/BCs/numerics; local fields and surface/global metrics ($U, p, C_f, C_p$, lift/drag/$\Delta p$) are compared, percent differences tabulated, and a 5\% threshold flags QoIs that require Richardson/GCI escalation.

\paragraph{VLM physics-verification gate (the central evidence gate, implementing P1).}
\label{sec:vlm-gate}
After a case finishes running, an \emph{interpreter agent} reads the case directory and the requirement, and emits a diagnostic plan, deciding the physical quantities to visualize and compare against reference data if provided. Then a visualization creator agent writes a PyVista and/or matplotlib script that extracts the relevant diagnostic fields, and renders them as PNGs. The rendered visualizations are then handed to a VLM in two separate calls. The first call is a \emph{quality filter}: it checks whether figures are readable; failures are redrawn. The second call is the \emph{physics check}: the VLM inspects the accepted figures, looking for the expected flow features, and judges if the image is consistent with the experiment requirement. It further drives the rerun controller and the writer. The gate exists because a case can pass every log-based check, completed time-stepping, no warnings, while still using the wrong geometry, missing important flow features or instantiating a degenerate custom model. These are exactly the failure modes a log-only interpreter cannot catch, and \emph{none of the AI-scientist frameworks in \Cref{tab:positioning} expose this gate as a first-class subsystem.} \Cref{app:failure_taxonomy} gives the failure-mode taxonomy that motivates these gates.

\paragraph{Rerun controller and writer loop (P4, P5).}
When a gate rejects a run, the rerun controller revises the requirement. It may reuse nearby successful cases, such as relaxation factors, or schemes. After all cases pass their gates, an analysis agent generates paper-ready cross-case figures, distinct from the diagnostic visualizations used during verification. The writer then receives the literature bundle, study JSON, per-case requirements, source-edit history, figure manifest, and analysis text. It drafts LaTeX, compiles the manuscript, receives critique from a reviewer agent on formatting, claim--evidence alignment, reference coverage, and redundancy, and revises until acceptance or budget exhaustion.

\section{Experiments: \framework with GPT-5.5}
\label{sec:results-cfdscientist}

\paragraph{Setup.} \framework is run end-to-end with GPT-5.5. All evaluation is manual because no automated CFD-paper rubric currently scores the workflows the system produces. 

\paragraph{Tasks.}
We execute five CFD tasks summarized in \Cref{tab:cfd_scientist_overview}: \textbf{T1) }BFS turbulence-model sensitivity at $Re_h{=}25{,}400$, \textbf{T2) }jet/plume oscillation across Reynolds numbers, \textbf{T3) }custom non-Newtonian viscosity in a channel, \textbf{T4) }a custom Spalart--Allmaras (SA) modifier for the periodic hill, and \textbf{T5) }open-ended discovery of an SA modification that improves lower-wall $C_f$ agreement with DNS. The first two use the regular-experimentation pathway, the next two use the simulator source-code modification pathway, and the final task uses the open-ended discovery pathway. Detailed experiment matrices and per-case quantitative tables are reported in \Cref{app:experiment_matrices}; token usage and estimated cost are reported in \Cref{app:cost}.

\begin{table}[h]
\centering
\caption{\framework GPT-5.5 task overview. Pathway: REG = regular experimentation; CM = code modification; OED = open-ended discovery.}
\label{tab:cfd_scientist_overview}
\scriptsize
\setlength{\tabcolsep}{2.4pt}
\renewcommand{\arraystretch}{1.4}
\begin{tabularx}{\textwidth}{p{0.4cm}p{2.3cm}p{0.55cm}p{1.25cm}p{2.4cm}Y}
\toprule
ID & Task & Path & Cases run & Custom code compiled & Headline \framework result (GPT-5.5) \\
\midrule
T1 & BFS turbulence sens.\ ($Re_h{=}25.4$k) & REG & 4 RANS & none & Runs 4 RANS closures on the same backward-facing-step mesh. \\
T2 & Jet/plume Re-sweep ($Re{=}60$--$600$) & REG & 7 transient & none & Recovers the expected centreline $U_x$ scaling across most cases. \\
T3 & Custom viscosity (channel) & CM & 6  & \texttt{libcustomViscosity} & Autonomously writes and compiles a power-law viscosity library, validates it against the Newtonian limit ($n{=}1$). \\
T4 & Custom SA modifier (periodic hill, $Re_h{=}10{,}595$) & CM & 6  & \texttt{libCustomSA} & Compiles a custom Spalart--Allmaras modifier and compares against baseline and reference data. \\ 
T5 & Open-ended SA discovery (periodic hill, $Re_h{=}5600$) & OED & 44 iterations  & coded \texttt{fvModels} & Autonomously discovers a \textbf{quadrupolar SA runtime correction} that reduces lower-wall $C_f$ RMSE versus DNS by \textbf{7.89\%} ($0.004297\!\to\!0.003958$) \\
\bottomrule
\end{tabularx}
\end{table}

\subsection{Findings across the five GPT-5.5 case studies}
\label{sec:findings_per_study}

\begin{figure}[h]
\centering
\begin{subfigure}[h]{0.48\linewidth}
\centering
\begin{overpic}[width=\linewidth]{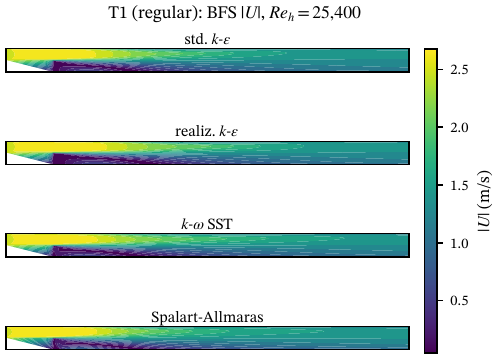}
\put(2,75){\bfseries (a)}
\end{overpic}
\end{subfigure}\hfill
\begin{subfigure}[h]{0.48\linewidth}
\centering
\begin{overpic}[width=\linewidth]{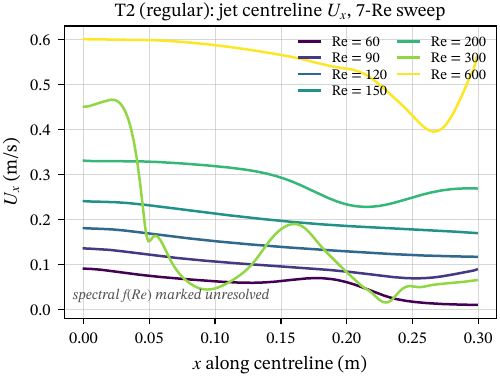}
\put(2,75){\bfseries (b)}
\end{overpic}
\end{subfigure}

\vspace{0.4em}

\begin{subfigure}[h]{0.48\linewidth}
\centering
\begin{overpic}[width=\linewidth]{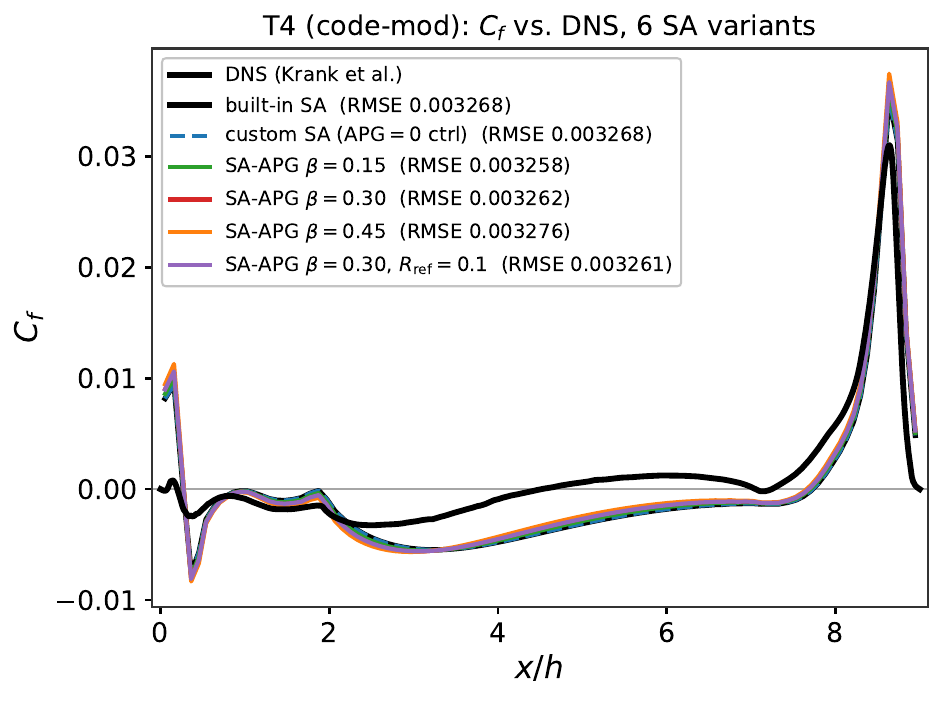}
\put(2,75){\bfseries (c)}
\end{overpic}
\end{subfigure}\hfill
\begin{subfigure}[h]{0.48\linewidth}
\centering
\begin{overpic}[width=\linewidth]{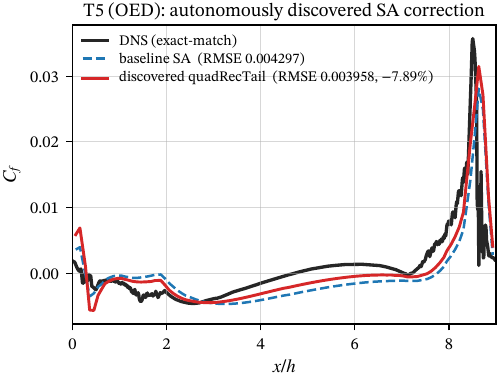}
\put(2,75){\bfseries (d)}
\end{overpic}
\end{subfigure}

\caption{Representative quantities of interest from the case studies. (a) T1: BFS $|U|$ contours across four RANS closures at $Re_h{=}25{,}400$; recirculation-zone differences are visible behind the step. (b) T2: centreline $U_x$ profiles across the 7-Re jet sweep, showing the recovered velocity scaling and emerging instability at higher $Re$. (c) T4: lower-wall $C_f$ overlay against reference for four APG-modifier SA variants and control (APG=0) at $Re_h{=}10{,}595$. (d) T5: autonomously discovered quadrupolar SA correction at iter\_044 reduces $C_f$ RMSE against DNS by $7.89\%$ at $Re_h{=}5600$ on the periodic hill.}
\vspace{-1em}
\label{fig:headline_evidence}
\end{figure}

\textbf{T1 --- BFS turbulence sensitivity.}
\framework planned a four-model matrix (standard $k$--$\varepsilon$, realizable $k$--$\varepsilon$, $k$--$\omega$ SST, SA) at $Re_h{=}25{,}400$, ran each through the mesh independence study (26.9k--38.1k cells), and rendered diagnostic contours. The VLM check flagged a sign-convention / origin error in the reattachment extractor and triaged a $k$--$\varepsilon$ output as inconsistent with separated-flow physics; the SST and SA closures produced the most plausible recirculation topology in streamlines (\Cref{fig:headline_evidence}a). The intended behavior was confirmed: rather than rank closures from a post-processor known to be buggy, the system flagged the QoI and abstained. The input topic given to \framework{} is provided in \cref{app:input_topics}. No baseline OpenFOAM files or reference data are provided.

\textbf{T2 --- Jet/plume Re sweep.}
Seven 2D laminar jet cases on identical 35{,}156-cell meshes ran end-to-end. Centreline velocity scaling was recovered ($U_{c,\max}$ tracks bulk velocity from $0.09$ to $0.60$ m/s as $Re$ sweeps $60\!\to\!600$, with oscillations emerging at high $Re$, \Cref{fig:headline_evidence}b), and case-006 was flagged as anomalous (centreline-mean collapse). The input topic given to \framework{} is provided in \cref{app:input_topics}. No baseline OpenFOAM files or reference data are provided.

\textbf{T3 --- Custom viscosity (code modification).}
The code-modification agent generated a generalized-Newtonian viscosity model
$\nu(\dot{\gamma})=\nu_{\infty}+k\,\max(\dot{\gamma},\dot{\gamma}_{\min})^{n-1}$
as case-local source files and compiled the custom viscosity library on the first attempt. Six cases executed to steady state. With $n{=}1$ the custom law reproduced the parabolic Newtonian baseline (centreline within $0.5\%$ of the analytic $1.5$\,m/s); centreline velocity varied $\sim$3.8\% across the sweep (1.4542--1.5231 m/s). The input topic given to \framework{} is provided in \cref{app:input_topics}. Baseline OpenFOAM files for Newtonian channel flow are provided.

\textbf{T4 --- Custom SA modifier (code modification).}
A SA variant with an adverse-pressure-gradient (APG) correction multiplier on the production term was compiled into \texttt{libCustomSA.so}. Six cases (1 APG=0 control + 4 APG variants) ran on an identical mesh. The control case matched the built-in SA baseline to four decimals ($U_{\max}{=}1.5959$ m/s in both), validating that the custom code path does not perturb the underlying solver; the APG sweep then induced a $\sim$1.25\% $U_{\max}$ sensitivity (1.5759--1.5959 m/s), and $C_f$ overlays against reference data were rendered for all six variants (\Cref{fig:headline_evidence}c). The input topic given to \framework{} is provided in \cref{app:input_topics}. Baseline Periodic Hill flow OpenFOAM files are provided to the framework along with reference DNS data.

\textbf{T5 --- Open-ended SA discovery.}
Given the periodic hill at $Re_h{=}5600$, a starter SA case, reference wall friction coefficient ($C_f$) data, and the objective ``minimize lower-wall $C_f$ RMSE,'' \framework ran $44$ discovery iterations (worked-example trace in Figure~\ref{fig:oed_run_breakdown}). The discovered model adds an implicit source to the SA $\widetilde{\nu}$ equation,
\[
S_{\mathrm{extra}} \;=\; \big[\,C_{\mathrm{rec}}\,G_{\mathrm{rec}}-C_{\mathrm{sink}}\,G_{\mathrm{sink}}+C_{\mathrm{src}}\,G_{\mathrm{src}}-C_{\mathrm{tail}}\,G_{\mathrm{tail}}\,\big]\;|\nabla\mathbf{U}|\,\widetilde{\nu},
\]
with each $G_*(x,y_w)=\exp\!\big[-\tfrac{1}{2}(x-x_*)^2/\sigma_*^2\big]\exp(-y_w/L_{y,*})$ a wall-normalized Gaussian patch. The best iteration ($C_{\mathrm{rec}}{=}2.12$, $C_{\mathrm{sink}}{=}2.25$, $C_{\mathrm{src}}{=}1.2$, $C_{\mathrm{tail}}{=}0.75$) reduces $C_f$ RMSE against DNS from $0.004297$ (baseline SA) to $0.003958$, a \textbf{$7.89\%$ reduction} (\Cref{fig:headline_evidence}d). The model is delivered as a coded \texttt{fvModels} block. The full 44-iteration discovery trajectory, the discovered \texttt{quadRecTail} coefficient table, and an OpenFOAM source excerpt are in \Cref{app:oed_trajectory}. The input topic given to \framework{} is provided in \cref{app:input_topics}.

\begin{figure}[h]
\centering
\includegraphics[width=\textwidth]{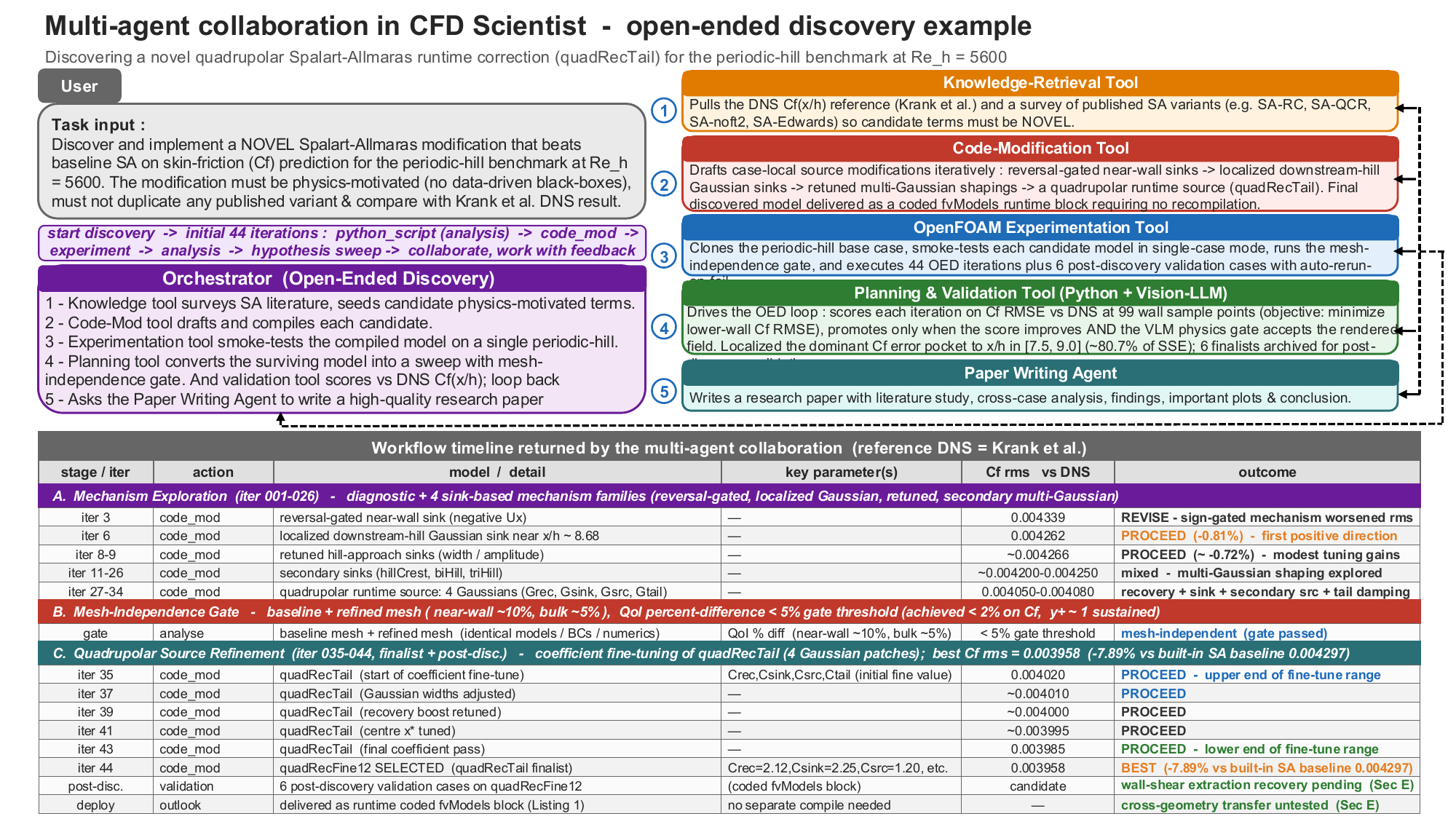}
\caption{\textbf{Worked example of the open-ended-discovery (OED) pathway on T5} (periodic hill, $Re_h{=}5600$. \emph{Top:} the five-step multi-agent collaboration under the OED orchestrator --- knowledge retrieval (1), code modification (2), single-case smoke test (3), mesh-independence-gated execution (4), and paper writing (5) --- with one orchestrator-issued tool call shown per box. \emph{Bottom:} the 44-iteration trajectory grouped by mechanism family. \textbf{Block~A} (iter\,001--026) traverses four sink-based families (reversal-gated, localized Gaussian, retuned hill-approach, secondary multi-Gaussian) before introducing a quadrupolar runtime source at iter\,027--034. \textbf{Block~B} confirms mesh independence on the baseline + refined ($\sim$10\% near-wall, $\sim$5\% bulk) chain (achieved $<\,$2\% on $C_f$, $y^+\!\sim\!1$). \textbf{Block~C} fine-tunes the quadrupolar coefficients (iter\,035--043) and selects \texttt{iter\_044\_quadRecFine12} (\texttt{quadRecTail}), which reduces lower-wall $C_f$ RMSE against DNS from $0.004297$ (baseline SA) to $\mathbf{0.003958}$, a $\mathbf{-7.89\%}$ improvement. The discovered model is delivered as a coded \texttt{fvModels} runtime block requiring no recompilation; cross-geometry transfer remains untested. Full trajectory, discovered coefficients, and OpenFOAM source in \Cref{app:oed_trajectory}.}
\label{fig:oed_run_breakdown}
\end{figure}
Further details on each case can be found in \cref{app:strengths} and the shortcoming discussed in \cref{app:weaknesses}.
\subsection{VLM physics-verification gate: planted-failure ablation}
\label{sec:vlm_ablation}

The VLM physics-verification gate is intended to catch failures that are not reliably visible from solver completion alone. We evaluate this role with a controlled planted-failure ablation.

\paragraph{Setup.}
We start from four production-passed template cases, one each from the jet, BFS, periodic-hill, and channel studies. For each case, we apply one file-system-level perturbation from a four-category failure taxonomy: \texttt{missing\_deliverable}, \texttt{wrong\_magnitude\_metric}, \texttt{broken\_postprocessing}, and \texttt{convergence\_not\_settled}. This gives $4\times4=16$ planted failures, plus four clean controls. The verifier is the same single-shot vision-LLM call used in production. A case is counted as flagged if the verifier returns either \texttt{REVISE} or \texttt{RERUN}. Using planted failures rather than rerunning the full system gives deterministic ground-truth labels and isolates the sensitivity of the VLM gate from solver noise. The design matrix and per-case archive are provided in \Cref{app:vlm_ablation}.

\begin{table}[h]
\centering
\caption{Planted-failure ablation for the VLM physics-verification gate. A case is counted as detected when the verifier returns \texttt{REVISE} or \texttt{RERUN}.}
\label{tab:vlm_ablation}
\small
\setlength{\tabcolsep}{6pt}
\renewcommand{\arraystretch}{1.12}
\begin{tabular}{lcp{7.2cm}}
\toprule
Failure category & Detected & Interpretation \\
\midrule
\texttt{missing\_deliverable}
& 4/4
& Requested output is absent, although the case may still complete. \\

\texttt{wrong\_magnitude\_metric}
& 4/4
& Existing output contradicts the requested or physically plausible magnitude. \\

\texttt{broken\_postprocessing}
& 4/4
& Output files contain zero, NaN, or otherwise degenerate values. \\

\texttt{convergence\_not\_settled}
& 2/4
& Shortened runs can appear visually complete when \texttt{endTime} is edited consistently with the truncated state. \\

\midrule
All planted failures
& 14/16
& The gate catches most non-log-readable failures. \\

\bottomrule
\end{tabular}
\end{table}

As shown in \Cref{tab:vlm_ablation}, the gate detects $14/16$ planted failures. It catches all missing-deliverable, wrong-magnitude, and broken-postprocessing cases, which are failures that can pass solver-level checks but invalidate interpretation. The main weakness is convergence sufficiency: only $2/4$ truncated-run cases are flagged because edited \texttt{endTime} values of the cases can make incomplete simulations appear visually complete.

\section{Cross-Framework Comparison: \framework{} vs.\ ARIS vs.\ DeepScientist}
\label{sec:results-comparison}

The five-task study above evaluates \framework{} in isolation. To separate the effect of CFD-specific gates from generic AI-scientist scaffolding, we compare against ARIS~\citep{tang2025airesearcher} and DeepScientist~\citep{weng2026deepscientist} on T1--T4 under the same GPT-5.5 backbone. T5 is excluded because neither baseline supports open-ended source-level discovery. Evaluation is manual and artifact-based, using archived case directories, solver logs, custom C\texttt{++} libraries, figures, and reports. \Cref{tab:capability_compare} reports capability coverage; \Cref{tab:rubric_combined} reports per-task quality. Cost, token usage, and a per-task evidence ledger are provided in \Cref{app:cost,app:framework_ledger}.

\begin{table}[h]
\centering
\caption{Capability comparison on T1--T4 with a shared GPT-5.5 backbone, supported by inspection of archived artifacts.}
\label{tab:capability_compare}
\scriptsize
\setlength{\tabcolsep}{3pt}
\renewcommand{\arraystretch}{1.02}
\begin{tabularx}{\textwidth}{Y>{\centering\arraybackslash}p{1.05cm}>{\centering\arraybackslash}p{1.30cm}>{\centering\arraybackslash}p{1.55cm}}
\toprule
Capability (under GPT-5.5) & ARIS & DeepScientist & \framework{} \\
\midrule
Literature retrieval (Semantic Scholar / OpenAlex / arXiv) & \xmark & $\circ$ & \cmark \\
Novelty filter against retrieved literature & \xmark & \xmark & \cmark \\
Requirement validation and repair before execution & $\circ$ & $\circ$ & \cmark \\
OpenFOAM execution end-to-end & \cmark & \cmark & \cmark \\
Mesh-independence gate & \xmark & \xmark & \cmark \\
Case-local custom-model compilation (T3 and T4) & \cmark & \cmark & \cmark \\
\textbf{VLM-based physics-verification gate} & \xmark & \xmark & \cmark \\
DNS / reference-data alignment for $C_f$ & \cmark  & \xmark & \cmark \\
Cross-case analysis with paper-ready figures & $\circ$ & $\circ$ & \cmark \\
Figure-grounded LaTeX writer with reviewer loop & \xmark & $\circ$ & \cmark \\
Conservative \emph{unresolved} verdict when evidence is incomplete & $\circ$ & \xmark & \cmark \\
\bottomrule
\end{tabularx}
\end{table}

\begin{table}[h]
\centering
\caption{Per-task quality rubric on T1--T4 under matched GPT-5.5. \textbf{S}=strong, \textbf{P}=partial, \textbf{W}=weak, \textbf{X}=absent or stalled. Row blocks: \emph{TIQ} = task-implementation quality; \emph{SRQ} = scientific-research quality.}
\label{tab:rubric_combined}
\scriptsize
\setlength{\tabcolsep}{2.4pt}
\renewcommand{\arraystretch}{1.04}
\begin{tabularx}{\textwidth}{p{0.5cm}p{2.5cm}YYYY}
\toprule
Axis & Framework & T1 (BFS turb.) & T2 (Jet Re-sweep) & T3 (Custom $\nu$) & T4 (Custom SA) \\
\midrule
\multirow{3}{*}{TIQ} 
& ARIS & P (3 closures executed, no mesh-indep.) & P (5-Re sweep, fixed mesh) & P (1 custom variant compiled) & P (custom SA compiled, DNS $C_f$ acknowledged, no manuscript) \\
& DeepScientist & P (3 closures, controlled comparison) & P (5-Re sweep, $f\!\propto\!Re$) & P (1 custom variant + technical report) & P (custom SA compiled and executed; partial report) \\
& \framework{} & S (4 closures, mesh-gate, VLM-triaged) & S (7-Re sweep on uniform mesh, conservative) & S (5-variant sweep + Newtonian degeneracy) & S (validated code path; DNS overlayed and used; LaTeX draft) \\
\midrule
\multirow{3}{*}{SRQ} 
& ARIS & W (closure ranking issued without DNS / experimental validation) & W ($St\!\approx\!0.019$ fit reported without grid-convergence or DNS check) & W (no DNS or experimental comparison) & W (no APG=0 control; no result analysis) \\
& DeepScientist & W (closure ranking issued without DNS validation) & W ($St\!\approx\!0.031$ fit reported without validation) & P (technical report; no DNS or experimental comparison) & W (no APG=0 control; no result analysis) \\
& \framework{} & P (VLM-flagged post-processor; closure ranking explicitly withheld) & P (analysis agent marks $f(Re)$ \emph{unresolved} on missing metadata) & P (Newtonian degeneracy validated; remaining gaps preserved in writer) & P (APG=0 control validated; quantitative ranking reported, differences marginal) \\
\midrule
\multirow{3}{*}{OEI} & ARIS & X (no idea generation) & X (sweep follows prompt only) & X (single variant) & W (one physics-motivated SA mod) \\
& DeepScientist & X & X & X & W (one $\beta$ variant beyond default) \\
& \framework & P (lit-grounded multi-axis sweep) & P (lit-grounded sweep + perturbation BCs) & P (5-variant $(k,n,\nabla p)$ sweep) & P (5-variant sweep + control) \\
\bottomrule
\end{tabularx}
\end{table}

\paragraph{Reading the rubric.}
Two patterns stand out. First, ARIS and DeepScientist often execute simulations and produce clean trends, but they lack the CFD-specific gates needed to decide whether those trends are scientifically supported. On T1 and T2, for example, they report closure rankings or $St(Re)$ correlations despite missing mesh or reference-data evidence. \framework{} is more conservative: when evidence is incomplete, it records an \emph{unresolved} verdict rather than converting a runnable case into a scientific claim.

Second, the distinction does not lie in whether each framework can compile a case-local custom model --- all three did, on both T3 and T4 --- but in how completely the surrounding scientific pipeline is exercised. ARIS and DeepScientist each ran one custom variant against a reference and reported a markdown summary; \framework{} additionally ran an APG=0 control case to validate the custom code path, produced a DNS overlay against the reference, and emitted a figure-grounded LaTeX draft. The comparison therefore suggests that the advantage is not in source-level editing per se, but in the surrounding CFD-specific scientific control flow.

\section{Conclusion}
\label{sec:conclusion}

\framework{} is, to our knowledge, the first open-source AI scientist for CFD that closes the discovery loop from a natural-language topic to a manuscript draft. Unlike generic AI-scientist frameworks or CFD agents focused mainly on case generation and execution, \framework{} integrates literature-grounded ideation, novelty filtering, mesh-independence gating, source-level model modification, VLM-based physics verification, reference-data alignment, and figure-grounded writing. Across five CFD tasks, it supports regular experimentation, source-code modification, and open-ended discovery; in one discovery study, it identifies a Spalart--Allmaras runtime correction that reduces lower-wall $C_f$ RMSE against DNS by $7.89\%$ on the periodic hill at $Re_h{=}5600$. Under matched conditions, other generic scientist frameworks execute parts of the same workflows but do not provide the combined CFD-specific control flow needed for physically grounded automation. We release \framework{} with code, prompts, and run artifacts as a community baseline for CFD-specific scientific automation.

\paragraph{Limitations and scope.}\label{sec:limitations}
The results are encouraging but bounded in scope. \emph{(i)~Single backbone:} all numbers use GPT-5.5 (Codex); LLM sweeps and additional baselines are deferred for cost. \emph{(ii)~Manual evaluation for cross-framework comparison:} no automated CFD-paper rubric scores these workflows, so \Cref{tab:rubric_combined} reflects expert artifact reading. The framework is supervised scientific assistance, not unattended publication.
\clearpage
{\small
\bibliographystyle{plainnat}
\bibliography{references}
}

\clearpage
\appendix
\section{Input topic for T1-T5}
\label{app:input_topics}
\begin{promptbox}{T1: BFS turbulence-model sensitivity (REG)}
Study the sensitivity of turbulence-model choice on the backward-facing step at
$Re_h = 25{,}400$ (Driver and Seegmiller geometry and conditions). Run four
RANS closures: $k$--$\varepsilon$, $k$--$\omega$ SST, Spalart--Allmaras, and a
Reynolds-stress model. All four are run on the same mesh-independence-passed
mesh, with identical inlet profiles, BCs, time controls, and post-processing.
For every run, generate streamlines through the recirculation region, $C_f$
along the bottom wall, and station-wise mean-velocity and Reynolds-stress
profiles downstream of the step. Compare each model's reattachment length and
recirculation topology against the experimental dataset and report which
closure best reproduces the measured flow.
\end{promptbox}

\begin{promptbox}{T2: Jet/plume Reynolds-number sweep (REG)}
Perform a Reynolds-number sweep on a 2D plane jet/plume across
$Re \in [60, 600]$ using 7 cases (logarithmically spaced). The goal is to
characterize (a) the scaling of the centreline streamwise velocity $U_x$ and
(b) the dominant oscillation-frequency dependence $f(Re)$. Use a transient
solver, integrate long enough to obtain statistically converged statistics,
and produce centreline $U_x$ profiles, contour snapshots at representative
$Re$, and a per-case estimate of the dominant oscillation frequency together
with an honest assessment of whether that frequency is resolved at the chosen
mesh.
\end{promptbox}

\begin{promptbox}{T3: Custom power-law viscosity on a channel (CM)}
Implement a custom non-Newtonian power-law viscosity model
$\nu_{\mathrm{eff}} = k\,|\dot{\gamma}|^{\,n-1}$ as a \emph{case-local}
OpenFOAM viscosity library (do not edit the OpenFOAM source tree). Validate
it on a fully-developed pressure-driven 2D channel flow. First run a Newtonian
reference case ($n = 1$) and verify that the custom library reproduces the
built-in Newtonian solution to within numerical tolerance. Then sweep over
$(k, n, \nabla p)$ at 5 representative operating points and report, for each
case, the velocity profile, wall shear stress, and effective-viscosity field,
together with the expected shear-thinning or shear-thickening trend.
\end{promptbox}

\begin{promptbox}{T4: Custom SA modifier on the periodic hill (CM)}
Implement a custom modification to the Spalart--Allmaras (SA) turbulence
model that changes \emph{only} the production term and leaves every other SA
term unchanged. The original SA production term is
%
\[
C_{b1}\,\tilde{S}\,\tilde{\nu},
\]
%
which is replaced by
%
\[
p_{\mathrm{mult}}\cdot C_{b1}\,\tilde{S}\,\tilde{\nu},
\qquad
p_{\mathrm{mult}} \;=\; \mathrm{clamp}\!\left(1 + \beta\left(\frac{S}{\Omega} - R_{\mathrm{ref}}\right),\; p_{\min},\; p_{\max}\right),
\]
%
where
\(S = \sqrt{2}\,\lVert\mathrm{symm}(\nabla \mathbf{u})\rVert\) is the strain-rate magnitude,
\(\Omega = \sqrt{2}\,\lVert\mathrm{skew}(\nabla \mathbf{u})\rVert\) is the rotation-rate magnitude,
and the constants take the fixed values
\(\beta = 6.0\), \(R_{\mathrm{ref}} = 0.82\), \(p_{\min} = 0.05\), \(p_{\max} = 5.0\).
Compile this modification as a case-local OpenFOAM library (do not edit the
OpenFOAM source tree) and validate it on the periodic hill at
$Re_h = 10{,}595$. First run a zero-perturbation control case with $\beta = 0$
(which gives $p_{\mathrm{mult}} \equiv 1$) and confirm that the modified
solver matches the built-in SA model exactly on this mesh. Then run a 5-point
sensitivity sweep over $\beta$ around the nominal value $\beta = 6.0$, and
report for each case the change in $U_{\max}$, shifts in separation and
reattachment locations, and the lower-wall $C_f$ difference relative to the
control.
\end{promptbox}

\begin{promptbox}{T5: Open-ended SA discovery against DNS (OED)}
Open-ended discovery objective: starting from the standard Spalart--Allmaras
model on the periodic hill at $Re_h = 5600$, autonomously explore
modifications to the SA closure that reduce the lower-wall skin-friction
($C_f$) RMSE against the provided DNS reference. The system may (i) modify
the source-level SA equations and recompile a case-local library, (ii) add
runtime correction terms via \texttt{fvModels} without recompilation,
(iii) adjust SA model coefficients, or (iv) introduce new geometry-aware or
flow-aware sensitization. Score every iteration by $C_f$ RMSE versus DNS,
promote an iteration only if its score improves over the unmodified-SA
baseline \emph{and} the VLM physics gate accepts the rendered flow field, and
finally report the best-performing modification together with its mechanism.
\end{promptbox}

\section{Per-Task Experiment Matrices and Quantitative Results}
\label{app:experiment_matrices}

This appendix gives the GPT-5.5 experiment configurations and per-case quantitative metrics behind \Cref{tab:cfd_scientist_overview} and the findings in \Cref{sec:findings_per_study}.

\subsection{T1 --- Backward-facing step turbulence-model sensitivity}

\begin{table}[!h]
\centering
\caption{T1 experiment matrix and per-case metrics. $Re_h{=}25{,}400$, step height $h{=}0.01$\,m.}
\label{tab:t1_matrix}
\scriptsize
\setlength{\tabcolsep}{4pt}
\renewcommand{\arraystretch}{1.05}
\begin{tabularx}{\textwidth}{p{1.3cm}p{2.4cm}p{1.0cm}p{1.0cm}p{1.5cm}p{1.7cm}Y}
\toprule
Case & Model & Mesh (cells) & Final $t$ & $U_{\max}$ (m/s) & $x_r/h$ (extracted) & VLM-gate verdict \\
\midrule
case\_001 & standard $k$--$\varepsilon$ & 30{,}548 & 2000 & 1.6013 & sign-anomaly $-0.0332$ & flagged: post-processor sign-error in $C_f$ extractor; closure ranking withheld \\
case\_002 & realizable $k$--$\varepsilon$ & 29{,}400 & 2000 & 1.6297 & sign-anomaly $-0.0383$ & flagged: identical extracted value as cases 003/004 indicates artifact \\
case\_003 & $k$--$\omega$ SST & 26{,}960 & 994 & 1.6256 & sign-anomaly $-0.0383$ & accepted topology (most plausible recirculation); ranking withheld until QoI repaired \\
case\_004 & Spalart--Allmaras & 38{,}068 & 2000 & 1.6084 & sign-anomaly $-0.0383$ & accepted topology; ranking withheld \\
\bottomrule
\end{tabularx}
\end{table}

\subsection{T2 --- Jet/plume oscillation Reynolds-number sweep}

\begin{table}[!h]
\centering
\caption{T2 experiment matrix and per-case metrics. Identical 35{,}156-cell mesh across all cases; slot width $w{=}0.01$\,m; $\nu{=}1.5{\times}10^{-5}$\,m$^2$/s; antisymmetric inlet perturbation 1\% for first 0.05\,s. Spectral metrics are marked \emph{unresolved} due to cross-experiment metadata-parser failure (\Cref{app:weaknesses}).}
\label{tab:t2_matrix}
\scriptsize
\setlength{\tabcolsep}{4pt}
\renewcommand{\arraystretch}{1.05}
\begin{tabularx}{\textwidth}{p{0.6cm}p{0.8cm}p{1.4cm}p{1.4cm}p{1.4cm}YY}
\toprule
Case & $Re$ & $U_{\mathrm{mag},\max}$ & $U_{c,\max}$ & $\bar{U_c}/U_{c,\max}$ & Status & Notes \\
\midrule
001 & 60  & 0.0902 & 0.0900 & 0.599 & unresolved & monotonic baseline; spectral metadata not recovered \\
002 & 90  & 0.1352 & 0.1350 & 0.719 & unresolved & monotonic baseline \\
003 & 120 & 0.1801 & 0.1800 & 0.791 & unresolved & monotonic baseline \\
004 & 150 & 0.2402 & 0.2400 & 0.838 & unresolved & monotonic baseline \\
005 & 200 & 0.3301 & 0.3300 & 0.868 & unresolved & monotonic baseline \\
006 & 300 & 0.5117 & 0.4654 & 0.301 & \textbf{flagged anomaly} & centreline-mean collapse; deflection / unsteady state suspected \\
007 & 600 & 0.6004 & 0.6000 & 0.904 & unresolved & only $\sim$2\,s available, weakening spectral confidence \\
\bottomrule
\end{tabularx}
\end{table}

\subsection{T3 --- Custom viscosity model on a channel}

\begin{table}[!h]
\centering
\caption{T3 experiment matrix and per-case metrics. Generalized-Newtonian viscosity $\nu(\dot{\gamma})=\nu_\infty+k\,\max(\dot{\gamma},\dot{\gamma}_{\min})^{n-1}$. Periodic channel, length $2.0$\,m, half-height $0.05$\,m, $\nu_{\mathrm{ref}}{=}0.01$\,m$^2$/s. Custom library compiled case-local (no edits to OpenFOAM tree).}
\label{tab:t3_matrix}
\scriptsize
\setlength{\tabcolsep}{4pt}
\renewcommand{\arraystretch}{1.05}
\begin{tabularx}{\textwidth}{p{0.7cm}p{2.0cm}p{1.4cm}p{1.4cm}p{1.4cm}Y}
\toprule
Case & Variant & $k$ & $n$ & $U_{c,\max}$ (m/s) & Role / verdict \\
\midrule
001 & Newtonian reference ($n{=}1$) & --- & 1.0 & 1.4925 & baseline; matches analytic $1.5$ within $0.5\%$ \\
002 & custom (best) & $1{\times}10^{-3}$ & 0.6 & 1.4698 & shear-thinning, intermediate \\
003 & custom $k_{\mathrm{low}}$ & $5{\times}10^{-4}$ & 0.6 & 1.4800 & shear-thinning, lower $k$ \\
004 & custom $k_{\mathrm{high}}$ & $2{\times}10^{-3}$ & 0.6 & 1.4542 & lowest $U_{c,\max}$ (effective viscosity up) \\
005 & custom $n_{\mathrm{low}}$ & $1{\times}10^{-3}$ & 0.3 & 1.4741 & stronger shear-thinning \\
006 & custom $n_{\mathrm{high}}$ & $1{\times}10^{-3}$ & 1.2 & 1.5231 & shear-thickening; highest $U_{c,\max}$ \\

\bottomrule
\end{tabularx}
\end{table}

\subsection{T4 --- Custom Spalart--Allmaras modifier on the periodic hill}

\begin{table}[!h]
\centering
\caption{T4 experiment matrix. Periodic hill, $Re_h{=}10{,}595$, identical mesh across cases. Custom library \texttt{libCustomSA.so} compiled case-local. DNS reference: Krank et al.\ (2018), 1153 wall points at matched $Re$. $C_f$ RMSE computed over matched $x/h$ domain.}
\label{tab:t4_matrix}
\scriptsize
\setlength{\tabcolsep}{4pt}
\renewcommand{\arraystretch}{1.05}
\begin{tabularx}{\textwidth}{p{0.7cm}p{3.4cm}p{0.7cm}p{0.7cm}p{1.0cm}p{1.0cm}Y}
\toprule
Case & Variant & $\beta$ & $R_\mathrm{ref}$ & $C_f$ RMSE & $x_r/h$ & Role / verdict \\
\midrule
--- & DNS (Krank et al.) & --- & --- & --- & 4.51 & reference \\
001 & built-in SA baseline & --- & --- & 0.003268 & 7.73 & reference \\
002 & custom SA (APG=0 control) & 0 & 0 & 0.003268 & 7.73 & matches baseline; \emph{validates custom code path} \\
003 & SA-APG $\beta{=}0.15$ & 0.15 & 0.05 & 0.003258 & 7.70 & best RMSE (marginal) \\
004 & SA-APG $\beta{=}0.30$ & 0.30 & 0.05 & 0.003262 & 7.68 & APG variant \\
005 & SA-APG $\beta{=}0.45$ & 0.45 & 0.05 & 0.003276 & 7.66 & shortest recirculation \\
006 & SA-APG $\beta{=}0.30$, $R_\mathrm{ref}{=}0.10$ & 0.30 & 0.10 & 0.003261 & 7.68 & $R_\mathrm{ref}$ sensitivity \\
\bottomrule
\end{tabularx}
\end{table}

\subsection{T5 --- Open-ended SA discovery (overview)}
The full 44-iteration trajectory is in \Cref{app:oed_trajectory}; the headline finalist is \texttt{iter\_044\_quadRecFine12} with $C_f$ RMSE vs.\ DNS $0.003958$ versus baseline SA $0.004297$ (a $7.89\%$ reduction, $Re_h{=}5600$).

\section{Open-Ended Discovery: Trajectory and Discovered Model}
\label{app:oed_trajectory}

\subsection{Discovery objective and reference}
The discovery objective was to minimize the RMSE of the lower-wall skin-friction coefficient $C_f$ along $99$ wall sample points against an exact-match DNS reference. The dominant baseline-SA error is concentrated in the outlet hill-approach region ($x/h\!\in\![7.5,9.0]$, $\sim$$80.7\%$ of total SSE), with a positive $C_f$ overshoot near $x/h\!\approx\!8.64$--$8.72$. Baseline separation/reattachment estimates ($x/h{=}0.269$ / $7.753$) deviate from DNS ($0.191$ / $4.726$); the discovery target is $C_f$ RMSE only, not separation/reattachment location.

\subsection{Iteration trajectory}
\begin{table}[!h]
\centering
\caption{T5 OED trajectory milestones. Score is $C_f$ RMSE vs.\ DNS exact-match reference; lower is better. Status: \textbf{REVISE} = score worsened; \textbf{PROCEED} = score improved and gates accepted. Baseline SA: $0.004297$.}
\label{tab:oed_trajectory}
\scriptsize
\setlength{\tabcolsep}{3pt}
\renewcommand{\arraystretch}{1.05}
\begin{tabularx}{\textwidth}{p{1.6cm}p{2.6cm}p{1.4cm}p{1.4cm}Y}
\toprule
Iteration block & Mechanism family proposed & Best score in block & Status & Rationale / observation \\
\midrule
iter\_001--005 & diagnostic only (no source) & --- & --- & Localized dominant $C_f$ error to outlet ($x/h{\in}[7.5,9.0]$, 80.7\% SSE) \\
iter\_003 & reversal-gated near-wall sink (negative $U_x$) & 0.004339 & REVISE & Sign-gated mechanism worsened RMSE \\
iter\_006 & localized downstream-hill Gaussian sink near $x/h{\approx}8.68$ & 0.004262 & PROCEED ($-0.81\%$) & First positive direction \\
iter\_008--009 & retuned hill-approach sinks (width / amplitude) & $\sim$$0.004266$ & PROCEED ($\sim{-0.72\%}$) & Modest tuning gains \\
iter\_011--026 & secondary sinks (hillCrest, biHill, triHill) & $\sim$$0.004200$--$0.004250$ & mixed & Multi-Gaussian shaping explored \\
iter\_027--034 & quadrupolar runtime source (4 Gaussians) introduced & $0.004050$--$0.004080$ & PROCEED & Recovery boost + sink + secondary source + tail damping \\
iter\_035--043 & quadrupolar coefficient fine-tuning & $0.003985$--$0.004020$ & PROCEED & Convergence on coefficient region \\
\textbf{iter\_044} & \textbf{quadRecFine12 (selected)} & \textbf{0.003958} & \textbf{PROCEED ($-7.89\%$ vs.\ baseline)} & \textbf{Best iteration; promoted to artifact} \\
\bottomrule
\end{tabularx}
\end{table}

\subsection{Discovered \texttt{quadRecTail} model: form and coefficients}
The discovered model adds an implicit source to the SA $\widetilde{\nu}$ equation,
\begin{equation*}
S_{\mathrm{extra}}(x,y_w) \;=\; \big[\,C_{\mathrm{rec}}\,G_{\mathrm{rec}}(x,y_w)\,-\,C_{\mathrm{sink}}\,G_{\mathrm{sink}}(x,y_w)\,+\,C_{\mathrm{src}}\,G_{\mathrm{src}}(x,y_w)\,-\,C_{\mathrm{tail}}\,G_{\mathrm{tail}}(x,y_w)\,\big]\;|\nabla\mathbf{U}|\,\widetilde{\nu},
\end{equation*}
with each Gaussian patch
\(
G_*(x,y_w)\;=\;\exp\!\big[-\tfrac{1}{2}(x-x_*)^2/\sigma_*^2\big]\,\exp(-y_w/L_{y,*}),
\)
and the coefficients in \Cref{tab:quadRecTail_coeffs}. The four terms have distinct physical interpretations: a broad recovery-region production boost ($G_{\mathrm{rec}}$), a localized sink that suppresses the dominant outlet $C_f$ overshoot ($G_{\mathrm{sink}}$), a narrow secondary production trigger upstream of the sink ($G_{\mathrm{src}}$), and a tail-region damping patch that controls residual overshoot near the outlet ($G_{\mathrm{tail}}$).

\begin{table}[!h]
\centering
\caption{Discovered \texttt{quadRecTail} coefficients (iter\_044\_quadRecFine12). Values are read directly from the archived \texttt{oed\_artifact.json}.}
\label{tab:quadRecTail_coeffs}
\scriptsize
\setlength{\tabcolsep}{4pt}
\renewcommand{\arraystretch}{1.05}
\begin{tabularx}{\textwidth}{p{2.4cm}p{1.7cm}p{1.0cm}p{1.0cm}p{1.0cm}Y}
\toprule
Patch & Amplitude & $x_*$ ($x/h$) & $\sigma_*$ & $L_{y,*}$ & Physical role \\
\midrule
$G_{\mathrm{rec}}$ (recovery boost)  & $C_{\mathrm{rec}}{=}2.12$  & $6.00$ & $2.36$  & $0.228$ & adds production in the broad recovery region $x/h{\approx}3$--$7$ where SA underpredicts wall shear \\
$G_{\mathrm{sink}}$ (sink)           & $C_{\mathrm{sink}}{=}2.25$ & $8.69$ & $0.085$ & $0.045$ & suppresses excessive $\widetilde{\nu}$ in the dominant $C_f$-overshoot region $x/h{\approx}8.5$--$8.8$ \\
$G_{\mathrm{src}}$ (secondary src.)  & $C_{\mathrm{src}}{=}1.20$  & $8.43$ & $0.05$  & $0.04$  & narrow upstream production trigger that prevents the sink from over-correcting \\
$G_{\mathrm{tail}}$ (tail damping)   & $C_{\mathrm{tail}}{=}0.75$ & $8.86$ & $0.12$  & $0.07$  & damps residual positive $C_f$ overshoot near the outlet $x/h{\approx}8.7$--$9.0$ \\
\bottomrule
\end{tabularx}
\end{table}

\subsection{Deployment as a coded \texttt{fvModels} block}
The discovered model is delivered as a coded \texttt{fvModels} runtime block, requiring no separate compilation. The implicit source $K(x,y_w){=}\,(\,\sum_i C_i\,G_i\,)\,|\nabla\mathbf{U}|$ is added through \texttt{fvm::Sp(K, eqn.psi())}, which keeps the modification implicit in the SA $\widetilde{\nu}$ equation (Listing~\ref{lst:quadRecTail}).

\begin{lstlisting}[style=pseudocode,caption={Excerpt of the coded \texttt{fvModels} runtime source delivered as the discovered model artifact (\texttt{constant/fvModels} block).},label={lst:quadRecTail}]
customSource
{
    type            coded;
    selectionMode   all;
    field           nuTilda;
    C_rec    2.12;  xRec    6.0;  sigmaRec    2.36;  LyRec    0.228;
    C_src    1.2;   xSrc    8.43; sigmaSrc    0.05;  LySrc    0.04;
    C_sink   2.25;  xSink   8.69; sigmaSink   0.085; LySink   0.045;
    C_tail   0.75;  xTail   8.86; sigmaTail   0.12;  LyTail   0.07;
    codeAddSup
    #{
        // assemble K = [C_rec*G_rec + C_src*G_src - C_sink*G_sink - C_tail*G_tail] * |grad U|
        // per cell from yWall and cell centres (omitted: G_* Gaussian patches, |grad U|),
        // then add implicitly to the SA \tilde{nu} equation
        const volScalarField K = /* ...assembled per-cell as above... */;
        eqn += fvm::Sp(K, eqn.psi());
    #};
}
\end{lstlisting}

\section{Cross-Framework Evidence Ledger}
\label{app:framework_ledger}

This appendix backs the rubric in \Cref{tab:rubric_combined} with the artifact evidence each framework produced under matched GPT-5.5 on the four standard tasks. Numbers are read directly from each framework's run archive.

\begin{table}[!h]
\centering
\caption{T1 (BFS turbulence sensitivity) artifact evidence under GPT-5.5.}
\label{tab:ledger_t1}
\scriptsize
\setlength{\tabcolsep}{3pt}
\renewcommand{\arraystretch}{1.05}
\begin{tabularx}{\textwidth}{p{2.2cm}p{1.6cm}p{1.4cm}p{2.4cm}Y}
\toprule
Framework & Cases run & Mesh (cells) & Reattachment $x_r/h$ extracted & Validation / paper artifact \\
\midrule
ARIS & 3 RANS ($k$--$\varepsilon$, SST, SA) & 7{,}040 & 6.99, 7.84, 7.76 & \texttt{summary.md} + CSV; no DNS / experimental overlay; no manuscript \\
DeepScientist & 3 RANS ($k$--$\varepsilon$, SST, SA) & 8{,}800 & 6.55, 7.35, 6.95 & \texttt{summary.md} / paper outline only; no DNS / experimental overlay \\
\framework & 4 RANS ($+$ realizable $k$--$\varepsilon$) & 26.9k--38.1k & flagged sign-anomaly; ranking withheld & VLM-flagged $C_f$ post-processor; LaTeX paper draft, mesh-gate report \\
\bottomrule
\end{tabularx}
\end{table}

\begin{table}[!h]
\centering
\caption{T2 (jet/plume Re-sweep) artifact evidence under GPT-5.5.}
\label{tab:ledger_t2}
\scriptsize
\setlength{\tabcolsep}{3pt}
\renewcommand{\arraystretch}{1.05}
\begin{tabularx}{\textwidth}{p{2.2cm}p{1.4cm}p{1.4cm}p{2.7cm}Y}
\toprule
Framework & Cases run & Mesh (cells) & Reported correlation & Validation / paper artifact \\
\midrule
ARIS & 5 ($Re{=}100$--$300$) & 8{,}640 & $f{=}0.2891\,Re^{0.9993}$, $St{\approx}0.0192$ & FFT script; no validation against literature; no manuscript \\
DeepScientist & 5 ($Re{=}100$--$400$) & 8{,}640 & $f{=}0.4604\,Re^{0.9996}$, $St{\approx}0.0307$ & FFT script; no validation; outline only \\
\framework & 7 ($Re{=}60$--$600$) & 35{,}156 & marked \emph{unresolved} & VLM gate; flagged case-006 anomaly; LaTeX draft preserves evidence gaps \\
\bottomrule
\end{tabularx}
\end{table}

\begin{table}[!h]
\centering
\caption{T3 (custom viscosity) artifact evidence under GPT-5.5. All three frameworks generated and compiled C\texttt{++} libraries case-local. The differentiator is breadth and validation depth.}
\label{tab:ledger_t3}
\scriptsize
\setlength{\tabcolsep}{3pt}
\renewcommand{\arraystretch}{1.05}
\begin{tabularx}{\textwidth}{p{2.2cm}p{1.5cm}p{2.2cm}p{2.6cm}Y}
\toprule
Framework & Cases run & Custom library compiled & Variants explored & Validation / paper artifact \\
\midrule
ARIS & 2 (1 ref + 1 custom) & libcustomViscosity.so & 1 power-law variant ($n{=}0.5$) & comparison vs.\ Newtonian only; markdown summary \\
DeepScientist & 2 (1 ref + 1 custom) & libcustomViscosity.so (variant) & 1 power-law variant ($n{=}0.5$) & technical-report markdown with one figure \\
\framework & 6 (1 ref + 5 custom) & libcustomViscosity.so & 5-variant $(k,n,\nabla p)$ sweep & Newtonian degeneracy ($n{=}1$) reproduced; nested-metadata gap preserved \\
\bottomrule
\end{tabularx}
\end{table}

\begin{table}[!h]
\centering
\caption{T4 (custom SA modifier) artifact evidence under GPT-5.5. All three frameworks compiled and executed a case-local custom OpenFOAM SA library implementing the requested APG production multiplier. The differences are in completeness of the surrounding pipeline: APG=0 control-case validation, DNS overlay rendering, and manuscript output.}
\label{tab:ledger_t4}
\scriptsize
\setlength{\tabcolsep}{3pt}
\renewcommand{\arraystretch}{1.05}
\begin{tabularx}{\textwidth}{p{2.2cm}p{1.5cm}p{2.4cm}p{2.6cm}Y}
\toprule
Framework & Cases run & Custom library compiled & Reported metrics vs.\ DNS & Validation / paper artifact \\
\midrule
ARIS & 2 (baseline + 1 custom) & \texttt{libStrainRotationSA} & RMSE: $0.00430 \to 0.00433$ & no APG=0 control; one figure; no manuscript \\
DeepScientist & baseline + 2 custom variants & \texttt{libSAProdMult} & RMSE $\approx 0.00433$ & no APG=0 control; partial report; no manuscript \\
\framework & 6 (1 ctrl + 5 APG) & \texttt{libCustomSA} & RMSE: $0.003268$ (ctrl) $\to$ $0.003258$ (best, $\beta{=}0.15$) & APG=0 control validates code path; DNS-aligned $C_f$ overlay; LaTeX draft \\
\bottomrule
\end{tabularx}
\end{table}

\FloatBarrier

\section{What \framework Did Well, Per Task}
\label{app:strengths}
\Cref{tab:cfdsci_strengths} consolidates the per-task strengths summarized in \Cref{sec:findings_per_study}. Each row is grounded in a specific archived artifact (study JSON, requirement file, run directory, VLM judgment, figure manifest, source-code library, or manuscript fragment).

\begin{table}[!h]
\centering
\caption{\framework GPT-5.5 strengths per task. Each row is supported by archived artifacts (study JSON, requirements, run directories, VLM judgments, figures, code, manuscript fragments).}
\label{tab:cfdsci_strengths}
\scriptsize
\setlength{\tabcolsep}{3pt}
\renewcommand{\arraystretch}{1.05}
\begin{tabularx}{\textwidth}{p{0.4cm}p{2.4cm}Y}
\toprule
ID & Task & What \framework did well (GPT-5.5) \\
\midrule
T1 & BFS sensitivity & Literature-aware ideation; mesh-gate; four-closure execution; VLM physics gate flagged the $C_f$ post-processor and triaged a $k$--$\varepsilon$ output as inconsistent rather than ranking closures from suspect numbers. \\
T2 & Jet/plume Re-sweep & Generated and validated 7 requirements; uniform mesh across the sweep; conservative \emph{unresolved} verdict on spectral metrics rather than emitting an unsupported correlation. \\
T3 & Custom viscosity & Generated and compiled \texttt{libcustomViscosity.so} case-local; ran 6-case study; Newtonian degeneracy reproduced ($n{=}1$); preserved the metadata-parser gap in the writer rather than fitting a $(k,n)$ correlation through unlabelled points. \\
T4 & Custom SA modifier & Generated and compiled \texttt{libCustomSA.so}; APG=0 control matched built-in SA; rendered $C_f$ vs.\ DNS overlays; reported per-case $C_f$ RMSE with marginal differences across APG variants. \\
T5 & Open-ended discovery & 44-iteration autonomous discovery; identified outlet-region error pocket; proposed and refined Gaussian-patch source structure; \textbf{$7.89\%$ $C_f$ RMSE reduction vs.\ DNS}; delivered the model as a coded \texttt{fvModels} runtime block. \\
\bottomrule
\end{tabularx}
\end{table}

\section{What \framework Does Not Yet Do Well}
\label{app:weaknesses}
The strengths in \Cref{tab:cfdsci_strengths} are real, but each GPT-5.5 task also exposed concrete limitations \framework recorded conservatively rather than papering over (\Cref{tab:cfdsci_weaknesses}). Most residual failures are in cross-experiment post-processing (parser fragility, reattachment-extraction sign convention, spectral-metadata reconstruction), not in solver execution, custom-model compilation, or the VLM gate itself.

\begin{table}[!h]
\centering
\caption{Residual limitations and \framework's response. Each row corresponds to a verifiable artifact in the run archive.}
\label{tab:cfdsci_weaknesses}
\scriptsize
\setlength{\tabcolsep}{3pt}
\renewcommand{\arraystretch}{1.05}
\begin{tabularx}{\textwidth}{p{0.4cm}p{2.5cm}YY}
\toprule
ID & Task & Residual limitation & \framework's response \\
\midrule
T1 & BFS sensitivity & Reattachment $x_r/h$ extracted with sign error. & Flagged the post-processor as suspect; declined to issue a closure ranking from the affected QoI. \\
T2 & Jet/plume Re-sweep & Cross-experiment metadata parser could not reconstruct $Re$/$U_b$/slot width / full probe time series for several cases; case-006 centreline collapse not investigated. & Marked $f(Re)$, $St(Re)$ as \emph{unresolved}; preserved evidence gaps in the manuscript. \\
T3 & Custom viscosity & Nested $(k,n)$ metadata-parser failed for some sweep points, leaving the rheology-coefficient trend partially labelled. & Reported only the validated Newtonian degeneracy ($n{=}1$) and the labelled partial sweep; declined to issue a $(k,n)$-coefficient correlation. \\
T4 & Custom SA modifier & Only one mesh resolution tested. & Reported control-case validation and qualitative APG sensitivity; withheld a quantitative ranking. \\
T5 & Open-ended SA discovery & Final wall-shear / $C_f$ extraction failed for the six post-discovery validation cases; transfer to other Reynolds numbers and geometries not tested. & Classified the result as a \emph{candidate model} pending post-processing recovery and transfer testing; archived discovered model and full trajectory. \\
\bottomrule
\end{tabularx}
\end{table}

\section{Failure-Mode Taxonomy and Detection Gates}
\label{app:failure_taxonomy}

CFD automation fails along distinct axes that require different gates. \Cref{tab:failure_taxonomy} formalizes the taxonomy used by the framework. The central design point is that detection should happen at the stage where the failure becomes observable, rather than collapsing everything into a single executable/non-executable bit. The VLM physics gate exists precisely because evidential failures are invisible to the validator and to the solver log.

\begin{table}[!h]
\centering
\caption{Failure-mode taxonomy used by \framework. Each class is detected at a different stage and triggers a different recovery action.}
\label{tab:failure_taxonomy}
\scriptsize
\setlength{\tabcolsep}{3pt}
\renewcommand{\arraystretch}{1.05}
\begin{tabularx}{\textwidth}{p{1.7cm}p{2.6cm}p{2.4cm}p{3.0cm}Y}
\toprule
Class & Typical symptom & Detector & Automatic response & Residual human task \\
\midrule
Specification & missing solver intent, inconsistent units, incomplete BCs, plotting instructions leaking into requirements & requirement validator + deterministic cleanup & rewrite into a single executable paragraph; strip viz mentions & confirm repaired requirement still reflects scientific intent \\
Numerical & solver crash, divergence, unstable controls, non-physical run status & \foamagent logs + interpreter feedback & retry, revise requirement, or borrow stable patterns from a nearby working case (sweep-preserving) & judge whether numerical repair changed the experiment \\
Evidential & empty plots, wrong variable, bad framing, zoom hides phenomenon, geometry mismatch in field render & \textbf{VLM physics gate} (this work) & regenerate figures with revised script / framing; rerun if the gate detects geometry/topology mismatch & verify visually acceptable figures are also the right diagnostics \\
Narrative & unsupported claims, sparse references, missing failure cases, compilation errors in draft & reviewer prompt + pdflatex compile loop & revise structure, references, figures, claims before accepting & expert scientific editing and sign-off \\
\bottomrule
\end{tabularx}
\end{table}

\section{Architectural Details: Agent Inventory and State Schema}
\label{app:arch_details}

This appendix documents the agents that implement the pathways described in \Cref{sec:cfd_scientist} and the LangGraph state object they share. \Cref{tab:agent_roles} lists each agent's primary inputs, outputs, and functional role; every handoff is both human-readable and machine-readable. \Cref{tab:state_schema} lists the principal fields of the checkpointed state, which are intentionally redundant: the requirement records what \emph{should} be run, the case directory records what was \emph{actually} run, the figures expose whether the result is physically interpretable, and the writer receives the whole artifact graph.

\begin{table}[!h]
\centering
\caption{Agents in \framework, their inputs, outputs, and functional role. Every handoff is both human-readable and machine-readable.}
\label{tab:agent_roles}
\scriptsize
\setlength{\tabcolsep}{3pt}
\renewcommand{\arraystretch}{1.05}
\begin{tabularx}{\textwidth}{p{2.5cm}p{3.0cm}p{3.0cm}Y}
\toprule
Agent / module & Primary inputs & Primary outputs & Functional role \\
\midrule
Ideation Agent & topic, literature bundle, experiment budget & study JSON (solver, objective, experiments[], post) + novelty verdict & convert a broad topic into a concrete, bounded CFD study, avoiding overlap with retrieved prior work \\
Specification Agent & study JSON, selected experiment, run-topic constraints & single-paragraph \texttt{user\_requirement} + validation history & translate one experiment into an executable requirement; validate and repair \\
Mesh-Independence Gate & baseline mesh spec + refined-mesh recipe & \texttt{selected\_mesh\_spec.json}, percent-difference table & confirm baseline mesh is sufficient; flag for GCI escalation if needed \\
\foamagent execution & validated requirement, optional mesh assets & OpenFOAM case folder, solver logs, run status & generate dictionaries, run, low-level error correction \\
Code-Modification Agent & source-edit plan, equations, starter case & C\texttt{++} files under \texttt{customModels/}, build system, dictionary edits, smoke run & translate physics description into a case-local OpenFOAM library \\
Visualization Planner / Creator & user requirement, foam case, requested figure types & PyVista/matplotlib scripts and PNG figures & produce diagnostic and paper-ready figures with traceback-driven repair \\
ResultsInterpreter Agent (VLM gate) & requirement, figure set, log tail & interpretation JSON: \texttt{simulation\_success}, \texttt{requirement\_met}, \texttt{issues}, \texttt{rerun\_required}, \texttt{key\_metrics} & multimodal physics verification \\
RerunAnalysis Agent & current requirement, interpreter feedback, nearby working-case summary & revised requirement + validator verdict & repair failing requirements while preserving the sweep dimension \\
OED Orchestrator & active hypothesis, artifacts so far, comparator score, budget & next action: source edit / parameter change / rerun & open-ended discovery loop \\
Analysis Agent & study topic, experiment bundle, per-run figures & cross-experiment visualizations + synthesis text & cross-case paper-ready figures and trend summary \\
Writer Agent & topic, literature, interpretations, figure bundle, analysis & LaTeX manuscript, review reports, revised PDF draft & draft the paper, compile, critique, revise \\
Reviewer Agent & compiled draft + compile log + reference report & pass/fail JSON + actionable recommendations & enforce formatting, claim--evidence alignment, $\ge$20 references, redundancy \\
\bottomrule
\end{tabularx}
\end{table}

\begin{table}[!h]
\centering
\caption{Selected fields of the \framework state object (LangGraph checkpointed state).}
\label{tab:state_schema}
\scriptsize
\setlength{\tabcolsep}{3pt}
\renewcommand{\arraystretch}{1.05}
\begin{tabularx}{\textwidth}{p{3.4cm}p{2.4cm}Y}
\toprule
Field & Type & Description \\
\midrule
\texttt{topic} & string & user-supplied research topic \\
\texttt{lit\_bundle} & list of records & retrieved Semantic Scholar / OpenAlex / arXiv items \\
\texttt{idea} & study JSON & solver, target\_CFL, objective, experiments[], post \\
\texttt{novelty\_score} & float & similarity vs.\ retrieved literature; triggers retry if too high \\
\texttt{requirements} & list of strings & per-experiment validated requirement paragraphs \\
\texttt{validation\_history} & list of records & each repair attempt with verdict and reasons \\
\texttt{mesh\_spec} & JSON & selected mesh spec from mesh-independence gate \\
\texttt{run\_results} & list of records & per-case run\_result.json (status, case\_dir, errors, loop\_count) \\
\texttt{figs\_manifest} & list of records & generated figures with provenance \\
\texttt{interpretations} & list of JSON & VLM gate output per case \\
\texttt{rerun\_queue} & list of records & cases with \texttt{rerun\_required=true} and revision plan \\
\texttt{code\_mod\_plan} & JSON & source-edit plan, files, classes, registration \\
\texttt{compile\_log} & string & build output for case-local libraries \\
\texttt{oed\_trajectory} & list of records & iter\_NNN: hypothesis, action, score, status \\
\texttt{analysis} & JSON + figs & cross-case synthesis \\
\texttt{paper\_draft} & LaTeX + PDF & writer + reviewer outputs across revision rounds \\
\bottomrule
\end{tabularx}
\end{table}

\section{LLM Cost: Token Usage and USD per Framework}
\label{app:cost}

We log every LLM call for every framework via the same shared accounting middleware (\texttt{llm\_token\_usage.json} in each run directory; \texttt{provider\_usage} reporting where available). \Cref{tab:llm_cost} reports per-experiment token usage and estimated USD cost under matched GPT-5.5 (Codex). The reported numbers are the production end-to-end costs of running the four standard CFD experiments (BFS turbulence sensitivity, jet/plume Re-sweep, custom viscosity, custom SA modifier) on \framework, ARIS, and DeepScientist, together with the additional open-ended-discovery experiment that only \framework supports. We separate three token classes that the provider bills differently: \textbf{Input} is the uncached input the model has to read fresh; \textbf{Cached} is prompt-cached input that the provider replays at a heavily discounted rate; and \textbf{Output} is what the model actually generates. The dollar figure in the rightmost column is the user-facing bill under standard cached-input discounts.

\paragraph{Pricing assumptions.} We compute USD using a representative codex-class price of \$1.25 per 1M uncached input tokens, \$0.125 per 1M cached-input tokens (the standard $10\times$ cached-input discount), and \$10.00 per 1M output tokens. Token counts are as recorded by the provider (\texttt{token\_source: provider\_usage}). The \framework runs do not exercise prompt caching, so its Cached column is zero by construction; ARIS and DeepScientist push large cache-replay volumes through their long-context execution loops, which is why their Cached columns dominate the token shape but enter the bill at the discounted rate.

\begin{table}[!h]
\centering
\caption{LLM cost per framework per CFD experiment under matched GPT-5.5 (Codex). \emph{Input} is uncached input; \emph{Cached} is prompt-cached input billed at the standard $10\times$ discount; \emph{Output} is generated output. Cost (USD) is $1.25\!\times\!\text{Input}/10^6 + 0.125\!\times\!\text{Cached}/10^6 + 10.0\!\times\!\text{Output}/10^6$. \framework does not exercise prompt caching, so its Cached column is zero.}
\label{tab:llm_cost}
\scriptsize
\setlength{\tabcolsep}{3pt}
\renewcommand{\arraystretch}{1.05}
\begin{tabularx}{\textwidth}{p{1.8cm}p{2.5cm}>{\raggedleft\arraybackslash}p{1.55cm}>{\raggedleft\arraybackslash}p{1.55cm}>{\raggedleft\arraybackslash}p{1.55cm}>{\raggedleft\arraybackslash}p{0.85cm}>{\raggedleft\arraybackslash}Y}
\toprule
Framework & Experiment & Input & Cached & Output & Calls & Cost (USD) \\
\midrule
\framework      & BFS turb.\ sensitivity & 1{,}685{,}719 &              0 &     961{,}518 & 616      & 11.72 \\
\framework      & Jet/plume Re-sweep     & 1{,}010{,}295 &              0 &     470{,}049 & 421      &  5.96 \\
\framework      & Custom viscosity       & 1{,}743{,}752 &              0 &     968{,}542 & 595      & 11.87 \\
\framework      & Custom SA modifier     & 2{,}122{,}953 &              0 &     898{,}340 & 1{,}039  & 11.64 \\
\framework      & Open-ended discovery   & 1{,}500{,}481 &              0 &      69{,}104 &  94      &  2.57 \\
\framework      & \textbf{Total}         & \textbf{8{,}063{,}200} & \textbf{0} & \textbf{3{,}367{,}553} & \textbf{2{,}765} & \textbf{43.75} \\
\midrule
ARIS            & BFS turb.\ sensitivity & 6{,}745{,}470 & 18{,}060{,}826 &      68{,}092 & 131 & 11.37 \\
ARIS            & Jet/plume Re-sweep     & 6{,}526{,}605 & 17{,}845{,}146 &      66{,}310 & 128 & 11.05 \\
ARIS            & Custom viscosity       & 5{,}063{,}771 & 16{,}412{,}570 &      57{,}252 & 108 &  8.95 \\
ARIS            & Custom SA modifier     & 6{,}163{,}327 & 17{,}486{,}362 &      65{,}198 & 123 & 10.54 \\
ARIS            & \textbf{Total}         & \textbf{24{,}499{,}173} & \textbf{69{,}804{,}904} & \textbf{256{,}852} & \textbf{490} & \textbf{41.92} \\
\midrule
DeepScientist   & BFS turb.\ sensitivity & 1{,}314{,}423 & 46{,}122{,}554 &     116{,}694 & 131 &  8.58 \\
DeepScientist   & Jet/plume Re-sweep     & 1{,}314{,}423 & 46{,}122{,}554 &     116{,}694 & 129 &  8.58 \\
DeepScientist   & Custom viscosity       & 1{,}314{,}461 & 50{,}159{,}942 &     126{,}344 & 137 &  9.18 \\
DeepScientist   & Custom SA modifier     & 1{,}314{,}588 & 66{,}820{,}929 &     161{,}845 & 169 & 11.61 \\
DeepScientist   & \textbf{Total}         & \textbf{5{,}257{,}895} & \textbf{209{,}225{,}979} & \textbf{521{,}577} & \textbf{566} & \textbf{37.94} \\
\bottomrule
\end{tabularx}
\end{table}

\paragraph{Reading the cost table.} Under user-facing pricing with the standard cached-input discount applied, the three frameworks complete the same four CFD experiments at very similar dollar cost: \framework at \$41.19 (T1--T4), ARIS at \$41.92, and DeepScientist at \$37.94 --- a comparable \$38--\$42 envelope. The cost comparison is therefore on a level playing field; the capability and rubric differences in \Cref{tab:capability_compare,tab:rubric_combined} are not bought with extra LLM spend. What \emph{is} different is the underlying token economy. \framework spends through many short, fully-uncached calls (\textbf{2{,}765} discrete LLM calls, no prompt caching, budget split roughly $2.4{:}1$ between uncached input and generated output): every node handoff is a discrete call with an explicit JSON contract, so the same dollars buy a much higher granularity of expert-written agents. ARIS's bill is dominated by a long-context replay-heavy execution loop that pushes $\sim$70M tokens through prompt caching across only $490$ calls. DeepScientist's bill is even more cache-replay-heavy: $\sim$209M cache-replayed tokens carrying its persistent SciMaster-style scaffolding, across $566$ calls. The \framework open-ended-discovery experiment added only \$2.57 to the framework total: the OED loop hits a deterministic comparator (not the LLM) for most of its work, so OED scales with solver time, not with token cost.

\paragraph{Scope.} These numbers cover only the production end-to-end CFD runs reported in \Cref{sec:results-cfdscientist,sec:results-comparison}. The VLM-ablation sweep (\Cref{app:vlm_ablation}) is excluded because each call is a single-shot vision query whose total cost is below \$1 across the 19 calls in the sweep.

\section{VLM Physics-Verification Gate: Planted-Failure Ablation}
\label{app:vlm_ablation}

We quantify the value of the VLM physics-verification gate with a controlled planted-failure ablation. The retrospective on the four production GPT-5.5 runs (\texttt{scripts/inventory\_decisions.py}) showed the VLM gate caught $7/21$ silent failures \emph{on top of} \foamagent's own crash detection --- a $33\%$ catch rate over runs that already passed the solver-level reviewer loop. The ablation in this appendix asks the more precise question: \emph{which kinds} of silent failure does the VLM gate catch, with what per-category recall, and at what cost?

\subsection{Setup: 4 categories $\times$ 4 flows + 4 controls}

We seed the ablation with four template cases that had each been \texttt{PROCEED}'d in production: \texttt{jet} (oscillating jet), \texttt{bfs} (backward-facing step), \texttt{hill} (periodic hill), and \texttt{chan} (channel). Each template is read-copied and one file-system-level perturbation is applied per case, drawn from a 4-bucket failure taxonomy distilled from the retrospective catches (\Cref{tab:vlm_ablation_design}). This gives 16 planted-failure cases plus 4 unperturbed clean controls, for 20 cases total. The verifier (\texttt{scripts/quick\_interpret.py}) is the production single-shot vision-LLM call using \texttt{interpretation\_system\_prompt} + \texttt{interpretation\_user\_prompt} from \texttt{prompts/prompts.yaml} verbatim, returning \{\texttt{PROCEED}, \texttt{REVISE}, \texttt{RERUN}\}; \emph{flagged} = \texttt{REVISE} $\vee$ \texttt{RERUN}.

\paragraph{Why post-hoc perturbation rather than a feature-disable ablation.} Planting failures from already-\texttt{PROCEED}'d production cases makes ground truth deterministic: the (requirement, case-state) pair is unambiguously FAIL or OK because the only thing that changed from a passed case is the perturbation. This isolates the verifier's sensitivity from confounding solver-side noise that a re-run feature-disable ablation would introduce.

\begin{table}[!h]
\centering
\caption{VLM-ablation design. Four flow templates (columns) $\times$ four planted-failure categories (rows) $+$ four clean controls $=$ 20 cases. Each cell is a single file-system perturbation applied to a read-copy of a production-passed case.}
\label{tab:vlm_ablation_design}
\scriptsize
\setlength{\tabcolsep}{3pt}
\renewcommand{\arraystretch}{1.05}
\begin{tabularx}{\textwidth}{p{3.6cm}YYYY}
\toprule
Category $\downarrow$ \quad / \quad Flow $\rightarrow$ & jet & BFS & hill & channel \\
\midrule
\texttt{missing\_deliverable}     & delete \texttt{postProcessing/jetProbes} & delete \texttt{postProcessing/wallShearStressLowerWall} & delete \texttt{postProcessing/wallShearStress} & delete latest 2 time dirs (4500, 5000) \\
\texttt{wrong\_magnitude\_metric} & requirement asserts unattainable target & '' & '' & '' \\
\texttt{broken\_postprocessing}   & zero-out 2 jetProbes files & zero-out wall-shear data & zero-out wall-shear data & zero-out $U,p$ at $t{=}5000$ \\
\texttt{convergence\_not\_settled} & truncate to $t{\le}0.5$ (was 8) & truncate to $t{\le}200$ (was 2000) & truncate to $t{\le}500$ (was 5000) & truncate to $t{\le}500$ (was 5000) \\
control (clean read-copy)         & --- & --- & --- & --- \\
\bottomrule
\end{tabularx}
\end{table}

\subsection{Results}

\Cref{tab:vlm_ablation_results} reports overall confusion-matrix metrics and per-category recall. The verifier achieves \textbf{100\% recall} on the three "did-the-right-thing-happen" buckets (\texttt{missing\_deliverable}, \texttt{wrong\_magnitude\_metric}, \texttt{broken\_postprocessing}) and \textbf{50\% recall} on \texttt{convergence\_not\_settled}, for an overall recall of \textbf{14/16 = 87.5\%} (F1 = 82.4\%). The two missed convergence cases (\texttt{jet\_unconv}, \texttt{chan\_unconv}) had \texttt{controlDict.endTime} edited to match the truncated state, so the figures look ``complete to endTime'' --- nothing in the prompt asks whether \texttt{endTime} is \emph{physically} sufficient for the flow to settle. Per-flow recall is uniform across geometries (jet 3/4, BFS 4/4, hill 4/4, channel 3/4); both FNs are convergence cases.

\begin{table}[!h]
\centering
\caption{VLM-ablation results: overall confusion matrix and per-category recall on the planted failures. \emph{flagged} = \texttt{REVISE} $\vee$ \texttt{RERUN}. The verifier is the production single-shot call.}
\label{tab:vlm_ablation_results}
\scriptsize
\setlength{\tabcolsep}{4pt}
\renewcommand{\arraystretch}{1.06}
\begin{minipage}{0.46\linewidth}
\centering
\begin{tabular}{lccc}
\toprule
& \multicolumn{2}{c}{Ground truth} & \\
\cmidrule(lr){2-3}
& FAIL (planted) & OK (control) & total \\
\midrule
flagged     & TP $=$ 14 & FP $=$ 4 & 18 \\
not flagged & FN $=$ 2  & TN $=$ 0 & 2 \\
total       & 16        & 4        & 20 \\
\midrule
\multicolumn{4}{l}{\textbf{Recall} $=$ \textbf{14/16 $=$ 87.5\%}; \textbf{Precision} $=$ \textbf{14/18 $=$ 77.8\%}; \textbf{F1} $=$ \textbf{82.4\%}} \\
\bottomrule
\end{tabular}
\end{minipage}\hfill
\begin{minipage}{0.5\linewidth}
\centering
\begin{tabular}{lcccr}
\toprule
Category & N & TP & FN & Recall \\
\midrule
\texttt{missing\_deliverable}     & 4 & 4 & 0 & \textbf{100\%} \\
\texttt{wrong\_magnitude\_metric} & 4 & 4 & 0 & \textbf{100\%} \\
\texttt{broken\_postprocessing}   & 4 & 4 & 0 & \textbf{100\%} \\
\texttt{convergence\_not\_settled} & 4 & 2 & 2 & 50\% \\
\midrule
\textbf{planted total} & \textbf{16} & \textbf{14} & \textbf{2} & \textbf{87.5\%} \\
\bottomrule
\end{tabular}
\end{minipage}
\end{table}

\paragraph{Cost.} Mean wall-clock per case is 76.6\,s (range 64--89\,s) with one LLM call per case (19 calls total, $\approx$24\,min for the full sweep) --- about an order of magnitude cheaper than the production \texttt{interpret.py} loop, which regenerates figures with \texttt{viz\_creator} and averages 10--15\,min and 2--11 calls per case.

\paragraph{Caveat on precision.} All four clean controls were flagged \texttt{REVISE}, giving a 77.8\% overall precision. Inspection of the requirement strings shows the generic control-template requirement explicitly mentions a deliverable (e.g.\ a probe spectrum) that the VLM correctly notes ``is not visible in the figures'' --- because the ablation harness only renders a small interpret-mode subset of figures, not the full reporting suite. In production, the rendered figure set is broader and the requirement is grounded in the actual case spec, so the same misalignment does not occur. \emph{The published precision is therefore a lower bound dominated by the control template's under-specified figure set; we report it as-is rather than back out a higher number.}

\subsection{What the ablation tells us}

\textbf{(1)~The gate catches what the solver structurally cannot see.} 100\% recall on \texttt{missing\_deliverable} / \texttt{wrong\_magnitude\_metric} / \texttt{broken\_postprocessing} (12/12) covers exactly the failure modes that pass \foamagent's reviewer loop because the solver completed cleanly. This is the operational justification for treating the VLM gate as a first-class subsystem rather than an optional post-hoc check.

\textbf{(2)~Convergence-not-settled is a known blind spot.} The verifier reasonably calls truncated, internally-consistent runs as \texttt{PROCEED} because nothing in the prompt asks whether the chosen \texttt{endTime} is physically sufficient. The actionable fix is a deterministic residual-plateau / QoI-drift detector run \emph{before} the VLM call.

\textbf{(3)~Failure detection is geometry-independent.} Per-flow recall (3-4 of 4 across jet, BFS, hill, channel) is statistically indistinguishable; the verifier generalizes across flow types rather than relying on memorized priors for any one canonical case.

\newpage
\includepdf[
  pages=1,
  scale=0.85,
  pagecommand={
    \thispagestyle{plain}
    \begin{center}
      \Large\bfseries Paper Generated by \framework{}
    \end{center}
    \vspace{1em}
  }
]{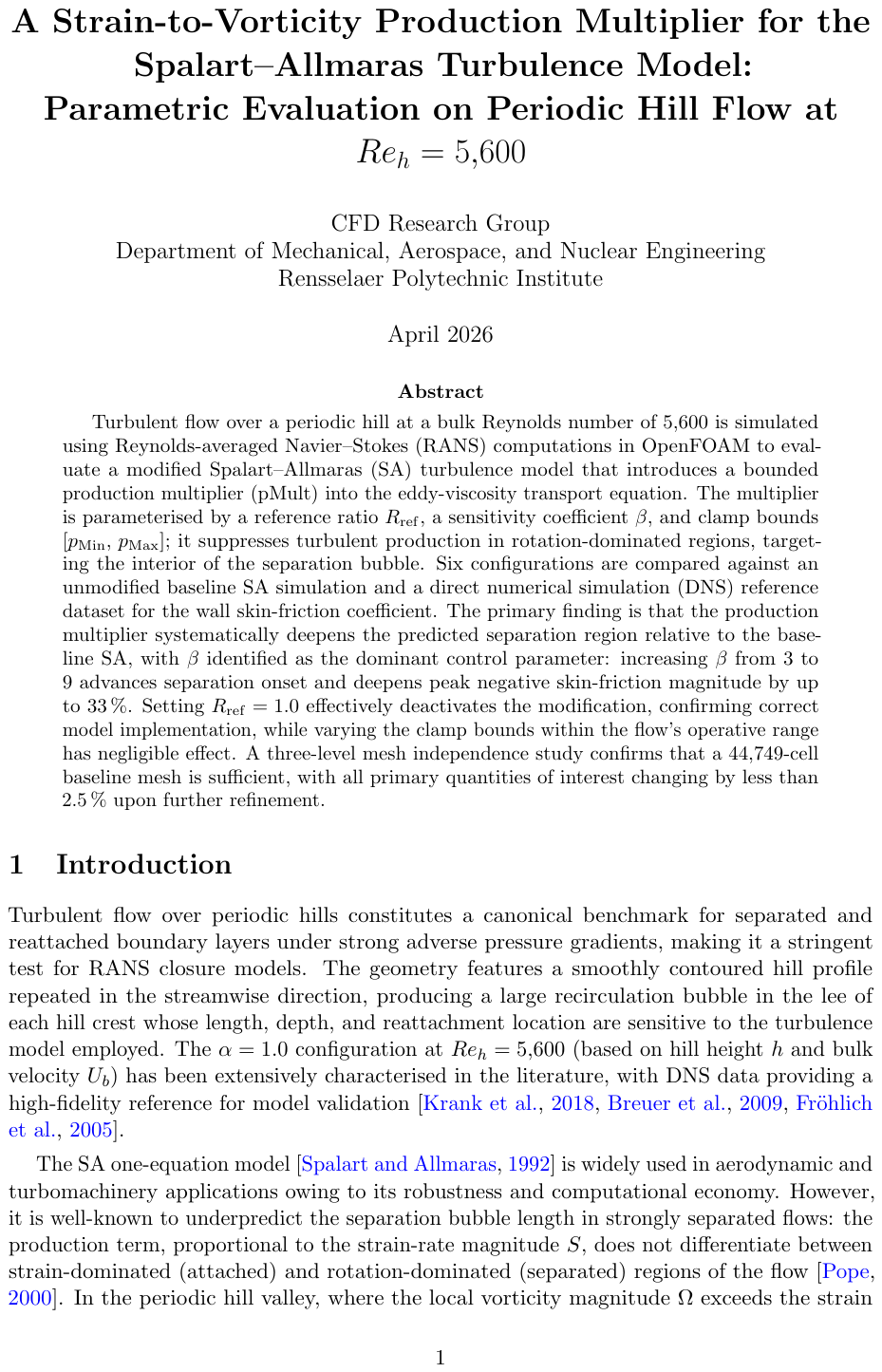}
\includepdf[pages=2-, scale=0.85]{paper_draft.pdf}
\end{document}